\documentclass[pra,reprint,showpacs,showkeys]{revtex4-1}
\usepackage{epsf,epsfig,amssymb,graphicx,times}
\usepackage{color}
\begin{document}

\title{Robustness of quantum key distribution with discrete and continuous variables to channel noise}

\author{Miko{\l}aj Lasota}
\author{Radim Filip}
\author{Vladyslav C. Usenko}
\email{Corresponding author. E-mail: 	usenko@optics.upol.cz}
\affiliation{Department of Optics, Palack\'{y} University, 17.\,listopadu 1192/12, 77146 Olomouc, Czech Republic}

\begin{abstract}
We study the robustness of quantum key distribution protocols using discrete or continuous variables to the channel noise. We introduce the model of such noise based on coupling of the signal
to a thermal reservoir, typical for continuous-variable quantum key distribution, to the discrete-variable case. Then we perform a comparison of the bounds on the tolerable channel noise between these two kinds of protocols using the same noise parametrization, in the case of implementation which is perfect otherwise. Obtained results show that continuous-variable protocols can exhibit similar robustness to the channel noise when the transmittance of the channel is relatively high. However, for strong loss discrete-variable protocols are superior and can overcome even the infinite-squeezing continuous-variable protocol while using limited nonclassical resources. The requirement on the probability of a single-photon production which would have to be fulfilled by a practical source of photons in order to demonstrate such superiority is feasible thanks to the recent rapid development in this field.
\end{abstract}

\maketitle

\section{Introduction}

Quantum key distribution (QKD) is the method of sharing a secret key between two trusted parties using nonclassical properties of quantum states. This enables the security of the key based on
physical principles contrary to the mathematical complexity in the classical cryptographic protocols. The first QKD protocols were suggested on the basis of single photons \cite{Bennett84} or entangled photon pairs \cite{Ekert1991} and, respectively, photon-counting measurements. The key bits were encoded to and obtained from the measurement of the states with the discrete spectrum and so the protocols were later referred to as discrete-variable (DV). Alternatively, schemes utilizing multiphoton quantum states of light and encoding the key using observables with the continuous spectrum \cite{Ralph99} were suggested on the basis of Gaussian modulation \cite{Weedbrook12} of squeezed \cite{Cerf2001} or coherent \cite{Grosshans2002,Weedbrook04} states and homodyne detection, and are referred to as continuous-variable (CV) QKD protocols. Both these families of protocols were successfully implemented \cite{Bennett92,*Muller95,*Jennewein00,*Naik00,*Tittel00,Grosshans2003,Lodewyck2007,*Huang2016,Madsen2012,Jouguet2013} and their security was analyzed with respect to individual \cite{Lutkenhaus96,*Slutsky98,*Bechmann06,Grosshans2003a}, collective \cite{Biham97,*Biham02,Navascues2006,*Garcia2006} or the most effective coherent attacks \cite{Kraus05,*Renner05,Leverrier2013}, also taking into account the effects of finite data ensemble size \cite{Hasegawa07,*Hayashi07,Scarani08,Leverrier2010,*Ruppert2014}. 

The applicability of all QKD protocols is limited by the imperfections of the devices used to prepare and measure quantum states and also by the properties of quantum channels, which are inclined to losses and noise \cite{Lutkenhaus99,*Brassard00,*Gottesman04,Filip2008,*Usenko2010a,*Jouguet2012,*Usenko2016,Garcia2009}. While it is important to understand which kind of protocols may be advantageous in specific conditions, at present there are no simple criteria for choosing either of their families for a particular task. The main reason for this is that making a fair comparison between DV and CV QKD protocols is hard due to the relativity of practical conditions and even different physical mechanisms leading to imperfections in the devices typically used in the protocol implementations. The only attempt to compare the performance of DV and CV systems done so far concerned the measurement-device independent systems and discussed practical conditions which can strongly vary depending on the wavelength, types of sources, channels and detectors being used, and set of optimistic or pessimistic assumptions being made about the possibility of an eavesdropper to attack the devices \cite{Xu15,*Pirandola15}. In our work we limit the discussion of realistic implementations of DV and CV schemes to a minimum. We mainly focus our attention on comparing the robustness of different types of protocols to the channel noise in the otherwise perfect set-ups. Later, we consider only finite nonclassical resources, \emph{i.e.} quality of single-photon states and finite amount of quadrature squeezing. 

Including the problem of channel noise in the QKD security analysis is more typical for the CV case. While it is well known that CV protocols can tolerate ideally any level of channel losses, the excess channel noise can be very harmful and even break the security of these protocols making QKD impossible. It can be considered as a main threat for their security. Indeed, the Gaussian excess noise, which is typically assumed in the CV QKD following the optimality of Gaussian collective attacks \cite{Navascues2006,*Garcia2006}, can break the security at the values below a shot-noise unit for a lossless channel and is further enforced by the channel losses \cite{Lodewyck2007,*Huang2016,Madsen2012,Jouguet2013}.

On the other hand, the analyses of DV QKD protocols performed so far usually focused on different setup imperfections, specifically multiphoton pulses and detection noise, which seem to be the main threats for security in this field. Even if various types of channel noise, originating \emph{e.g.}\,from  birefringence effect present in optical fibers, inhomogeneity of the atmosphere, changes of temperature or background light were sometimes included in these investigations \cite{Castelletto03,*Dong11}, they were usually described in a very simplified way, typically by using a single constant parameter, estimation of which could be made experimentally for a given, specific setup. This is especially true for the analyses of free-space DV QKD considering background light, arriving at Bob's detectors from other sources than the one used by Alice \cite{Miao05,*Bonato09,*Bourgoin13}. Disturbances of the states of photons traveling through a given quantum channel in the case of fiber-based QKD schemes were typically treated in a similar way, basing on the assumption that although these kind of effects can generally vary in time, the variations can be considered to be very slow comparing to the time needed for a single photon to propagate from Alice to Bob \cite{Dong11}. In this case, it is reasonable to assume that in short periods of time channel noise affects all of the traveling photons in the same way and can be described by a single, constant parameter. Such noise, called collective, was analyzed in many articles and a lot of possible countermeasures against it have been proposed, utilizing \emph{e.g.} Faraday mirrors \cite{Muller97,*Stucki02}, decoherence-free subspaces \cite{Zanardi97,*Kempe01,*Walton03,*Boileau04,*Li08}, quantum error-rejection codes \cite{Wang04a,*Wang04b,*Kalamidas05,*Chen06}, dense coding \cite{Cai04,*Wang05b,*Li09} or entanglement swapping \cite{Yang13}. However, no detailed analysis of the relationships between the transmittance of the channel connecting Alice and Bob, and the amount of tolerable channel noise has been presented so far and the influence of this relationship on the security of DV QKD protocols has never been analyzed, at least to our knowledge. At the same time due to the continuous improvement of realistic single-photon sources and detectors taking place nowadays, this issue gradually gains importance, especially since the links connecting Alice and Bob in commercial QKD applications may be more noisy than in the typical quantum-optical laboratories \cite{Eraerds10,Qi10}. 

In this paper we use in the DV protocols the model for excess channel noise basing on a typical model for CV QKD configuration. We analyze the security bound on such noise for both of these two cases under the assumption that Alice's sources and Bob's detection systems are perfect. Furthermore, we check the stability of the obtained results to the decreasing number of quantum signals exchanged by the trusted parties during the protocol in the finite-key regime. We also compare lower bounds on the secure key rate for the two schemes and find requirements for the nonclassicality of resources needed for their realistic implementations. For the case of ideal sources and detectors our study shows that while CV protocols can successfully compete with DV schemes when the channel transmittance is relatively high, the latter are superior than the former ones for long-distance channels. In this situation it turns out to be possible for DV protocols to beat infinite-squeezing CV schemes even when using realistic single-photon sources. The requirements on the quality of pulses produced by such sources, which would be needed in order to demonstrate this superiority in practice, turn out to be high but reachable by the current technology. For the thermal sources of noise with mean number of photons produced per pulse higher than $10^{-4}$ overcoming CV protocols with DV schemes can be possible only by using single-photon sources with at least $50\%$ probability of producing a non-empty pulse and negligible probability of multiphoton emission.

The paper is organized as follows. In Sec.\,\ref{Sec:Models} we describe the models for excess channel noise used in our analysis: first a standard model for CV QKD case and subsequently the analogous model for DV QKD case. We also derive there all the necessary formulae needed for assessing the security of Gaussian squeezed-state, BB84 and six-state protocols. Next, in Sec.\,\ref{Sec:NumericalResults} we numerically compare the maximal secure values of the channel noise that these protocols can tolerate on the transmittance of the channel connecting Alice and Bob in the case of perfect source and detection system. The comparison is done both in the asymptotic and the finite-key regimes. We also present analytical expressions approximating the maximal tolerable channel noise for DV and CV protocols in the limit of very low transmittance. More realistic situation is analyzed in Sec.\,\ref{Sec:RealisticCase}, where we investigate how the quality of Alice's source may influence the security of our models. Finally, Sec.\,\ref{Sec:Conclusions} concludes our work.

\section{Models for the channel noise in QKD}
\label{Sec:Models}

To assess the security of DV and CV QKD protocols we estimate the lower bound on the secure key rate per one pulse emitted by Alice's source. In the DV case this quantity can be expressed as \cite{Scarani09}
\begin{equation}
K^{(DV)}=p_{exp}\Delta I,
\label{eq:keyDV}
\end{equation}
where $p_{exp}$ denotes the probability for Bob to get a click in his detection system per pulse produced by the source and $\Delta I$ is the so-called secret fraction. Following the quantum generalization of the Csisz\'ar-K\"orner theorem \cite{Csiszar1978} performed by Devetak and Winter \cite{Devetak2005}, this quantity reads
\begin{equation}
\Delta I=\max[0,I_{AB}-\min\left\{I_{EA},I_{EB}\right\}],
\label{eq:DeltaImain}
\end{equation}
where $I_{AB}$ is the mutual information between Alice and Bob and $I_{EA}$ ($I_{EB}$) represents the amount of information Eve can gain on Alice's (Bob's) data upon an eavesdropping attack. On the other hand in the case of CV QKD protocols the lower bound on the secure key rate can be written simply as
\begin{equation}
K^{(CV)}=\Delta I,
\label{eq:keyCV}
\end{equation}
since, contrary to the DV case, all of the pulses emitted by Alice's source are registered by Bob's detection system in this situation. Generally speaking both the formulae  (\ref{eq:keyDV}) and (\ref{eq:keyCV}) should also contain the so-called sifting probability, representing the chance for the chosen settings of Bob's measurement setup to be compatible with a given signal sent by Alice. However, in the theoretical, asymptotic case, in which we assume that the key produced by Alice and Bob is infinitely long, its generation rate can be increased without compromising its security by performing highly asymmetric version of a given DV or CV protocol, making the sifting probability arbitrary close to one \cite{Scarani09,Lo05b}.

While the methods of calculating the lower bound on the secure key rate (\ref{eq:DeltaImain}) are substantially different in the DV and CV QKD, as we discuss in the following subsections, we develop the model of noise which can be applied to both these families of protocols using the same parametrization. The model is based on coupling every signal mode to an independent thermal reservoir with the coupling ratio $T$, which corresponds to the channel transmittance, and with the reservoirs being characterized by the mean number of thermal photons $\mu$ emitted per pulse. We study robustness of the DV and CV protocols to such thermal noise and derive and compare the security bounds in terms of the maximum tolerable mean numbers of noise photons.

\subsection{Channel noise in CV QKD}
\label{Sec:CVmodel}

CV QKD protocols typically use Gaussian states of light and respectively Gaussian modulation, which are compatible with the extremality of Gaussian states \cite{Wolf2006} and enable the security proofs against optimal Gaussian collective attacks \cite{Navascues2006,*Garcia2006}. In our study we consider the Gaussian squeezed-state protocol  based on the quadrature modulation and homodyne detection \cite{Cerf2001}. The reason why we choose this scheme instead of the more popular GG02 protocol \cite{Grosshans2002} is that the squeezed-state protocol is more resistant to the channel noise than GG02. This conclusion can be confirmed by comparing the results of our analysis performed for the squeezed-state protocol, presented in Sec.\,\ref{Sec:NumericalResults}, with the analogous results obtained for the GG02 scheme, shown in the Appendix \ref{Sec:GG02protocol}. Moreover, the squeezed-state protocol is the best known Gaussian CV QKD protocol in terms of the resistance to the channel noise \cite{Madsen2012}. Hence, demonstration of its inferiority to the DV protocols in that regard, shown in Sec.\,\ref{Sec:NumericalResults}, automatically implies that also other existing Gaussian CV QKD protocols cannot compete with the DV schemes.

\begin{figure}
\centering
\includegraphics[width=1.0\linewidth]{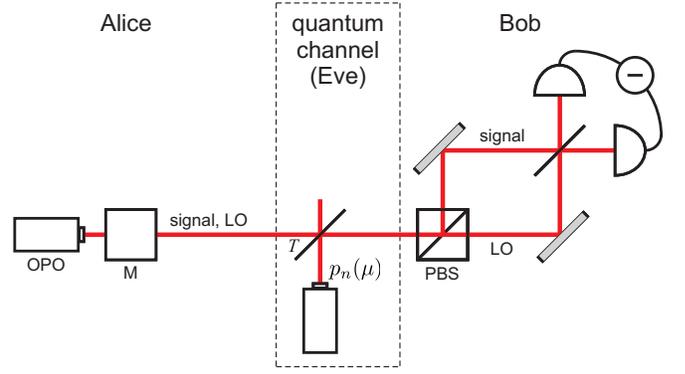}
\caption{(color online) CV QKD scheme with lossy and noisy quantum channel connecting Alice and Bob. The following abbreviations were used in this picture: OPO -- optical parametric oscillator, M -- amplitude/phase quadrature modulator, PBS -- polarization beam-splitter, LO - local oscillator.}
\label{fig:CVscheme}
\end{figure}

The squeezed-state protocol was shown to be secure against collective \cite{Navascues2006,*Garcia2006} and subsequently against general attacks \cite{Renner2009} in the asymptotic limit and against the collective attacks in the finite-size regime \cite{Leverrier2010}. In our analysis we assume that i) Alice uses a perfect source of quadrature-squeezed states with a quadrature variance $1/V \ll 1$ and that ii) Bob's homodyne detection is perfect with a unity efficiency and no uncontrollable noise. The scheme of the protocol, illustrated in Fig. \ref{fig:CVscheme}, is based on the squeezed signal state preparation by Alice using an optical parametric oscillator (OPO), phase/amplitude quadrature modulation based on the random Gaussian displacements applied in the modulator (M), and transmission along with the local oscillator (LO), being a phase reference for the homodyne measurement, through the Gaussian lossy and noisy channel. The remote party (Bob) splits the signal from the LO and performs homodyne measurement on the squeezed and modulated quadrature. The parties should swap between the bases (i.e. squeezing and modulating either of the two complementary quadratures) in order to perform the channel estimation, but in the following we assume that the channel estimation is perfect.

We now use Gaussian asymptotic security analysis to estimate the security bounds on the CV QKD protocols. To do so, following the Gaussian security proofs, we calculate the lower bound on the secure key rate in the reverse reconciliation scenario, which is known to be more robust against channel loss \cite{Grosshans2003} and being no less sensitive to the channel noise:
\begin{equation}
\label{LBCV}
K^{(CV)}=\max[0,I_{AB}-\chi_{BE}],
\end{equation}
where $I_{AB}$ is the mutual information shared between the trusted parties, and $\chi_{BE}$ is the Holevo bound \cite{Holevo2001}, upper limiting the information available to an eavesdropper from a collective attack in a given channel. To analyze the security of CV QKD we switch to the equivalent entanglement-based representation \cite{Grosshans2003a} so that Alice and Bob measure a two-mode entangled state shared between them through a quantum channel. The covariance matrix of the state is then given by
\begin{equation}
\label{gammaAB}
\gamma_{AB} =
\left( \begin{array}{cc}
V\mathbb{I} & \sqrt{T} \sqrt{V^2-1}\sigma_z \\
\sqrt{T} \sqrt{V^2-1}\sigma_z & [V T + (1 - T)W]\mathbb{I}
\end{array} \right),
\end{equation} 
where the diagonal matrix $\mathbb{I}=diag(1,1)$, $\sigma_z=diag(1,-1)$ is the Pauli matrix, $V$ is the variance of the modulated squeezed signal states, and $W=2\mu+1$ is the quadrature variance of the thermal noise state. The mutual information between the trusted parties then reads
\begin{equation}
\label{CVmutinf}
I_{AB}=\frac{1}{2}\log_2{\frac{V+W'}{1/V+W'}},
\end{equation}
where $W'=W(1-T)/T$. Following the pessimistic assumption that Eve is able to purify all the noise added to the signal, we estimate the Holevo bound as $\chi_{BE}=S(AB)-S(A|B)$ through the quantum (von Neumann) entropies $S(AB)$ derived from the symplectic eigenvalues \cite{Weedbrook12} $\lambda_{1,2}$ of the state described by the covariance matrix (\ref{gammaAB}), and $S(A|B)$ derived from the symplectic eigenvalue $\lambda_3$ of the state conditioned on Bob's measurement and described by the covariance matrix
\begin{equation}
\gamma_{A|B}=\gamma_A-\sigma_{AB}(X \gamma_B X)^{MP}\sigma_{AB}^T,
\end{equation}
where $\gamma_A=diag(V,V)$, $\gamma_B=diag([V T + (1 - T)W],[V T + (1 - T)W])$ are the matrices, describing the modes A and B individually; $\sigma_{AB}=\sqrt{T(V^2-1)}\sigma_z$ is the matrix, which characterizes correlations between the modes A and B, all being submatrices of (\ref{gammaAB}). MP stands for Moore-Penrose inverse of a matrix (also known as pseudoinverse applicable to the singular matrices), and $X=diag(1,0)$. Here with no loss of generality we assume that the x-quadrature is measured by Bob. Now the Holevo bound can be directly calculated as
\begin{equation}\label{holevo1}
\chi_{BE}=G\bigg(\frac{\lambda_1-1}{2}\bigg)+G\bigg(\frac{\lambda_2-1}{2}\bigg)-G\bigg(\frac{\lambda_3-1}{2}\bigg),
\end{equation}
which together with the mutual information (\ref{CVmutinf}) gives the lower bound on the secure key rate (\ref{LBCV}). Here $G(x)=(x+1)\log_2{(x+1)}-x\log_2x$ is the bosonic entropic function \cite{Serafini2005}. The bounds on the channel noise, characterized by $\mu$, are then derived by turning the secure key rate (\ref{LBCV}) to zero. 

We also consider the extension of the protocol, when trusted noise is added on the detection stage to improve the robustness of the protocol to the channel noise \cite{Garcia2009} (note that the heterodyne detection can be seen as the particular case of such noise addition and therefore was not considered in our study). This provides the maximum tolerable channel noise for a perfect CV QKD protocol with a given squeezing $1/V$ and upon given channel transmittance $T$.

\subsection{Channel noise in DV QKD}
\label{Sec:DVmodel}

\begin{figure}
\centering
\includegraphics[width=1.0\linewidth]{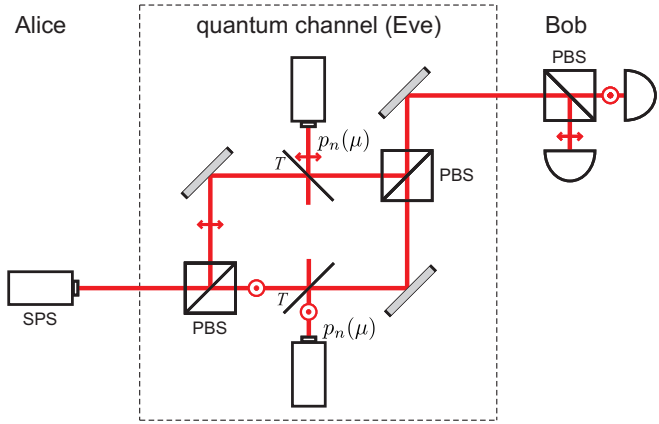}
\caption{(color online) Our model for DV QKD scheme with lossy and noisy quantum channel connecting Alice and Bob. The following abbreviations were used in this picture: SPS -- single-photon source, PBS -- polarization beam-splitter.}
\label{fig:DVscheme}
\end{figure}

Alternatively to the above-described CV scheme we consider the use of polarization-based BB84 \cite{Bennett84} and six-state \cite{Bruss98} protocols, both belonging to the family of DV protocols, to generate secure key by Alice and Bob. The scheme which we analyze is illustrated in  Fig.\,\ref{fig:DVscheme}. We assume that i) Alice's source is a perfect single-photon source and ii) Bob uses perfect single-photon detectors with no dark counts and unity efficiency. Our basic assumption on these detectors is that they do not have the ability to resolve the number of incoming photons. However, in the Appendix \ref{Sec:DifferentDetection} we analyze also the opposite possibility for the comparison. Since Alice's source is perfect, it never emits multiphoton pulses and Eve cannot perform photon-number-splitting attacks on the signal pulses. If so, Alice and Bob cannot gain anything by using decoy-pulse method \cite{Hwang03,*Wang05a,*Lo05} and we do not consider it in our analysis. 

In the model presented in Fig.\,\ref{fig:DVscheme} the channel noise, coupled to the signal during its propagation between Alice and Bob, is generated in two orthogonal polarizations by two independent sources of thermal noise. In fact this model is completely analogous to the one analyzed in Sec.\,\ref{Sec:CVmodel}, where two polarization modes are used to transmit the signal and the local oscillator. Since the effect of this noise on the bright local oscillator is negligible, it is not considered in the CV case. We assume here that the photons emitted by a given source of noise have the same polarization as signal photons transmitted through the channel to which it is coupled. We denote the probability of emitting $n$ noise photons by a given source by $p_n(\mu)$, where $\mu$ is the mean number of noise photons produced per pulse. For thermal noise this probability is given by
\begin{equation}
p_n(\mu)=\frac{\mu^n}{(\mu+1)^{n+1}}.
\end{equation}

Similarly to the CV case we assume that Eve fully controls the noise coupled to the signal in the quantum channel. Therefore, she can perform any attack which produces the same QBER as would be obtained by the trusted parties if there was no eavesdropper. We assume here that Eve executes the general collective attack, which is optimal for the DV QKD protocols under given QBER \cite{Kraus05,*Renner05}.

We also consider the possibility for Alice and Bob to perform so-called preprocessing \cite{Renner05}, allowing them to improve the security of the generated key by deliberately adding some noise to it before going to the stages of error correction and privacy amplification. This technique can be seen as the DV counterpart to the noise addition on the Bob's side considered in the CV case above in order to reduce the information which is available to Eve.

In the case without preprocessing, the most general collective attacks performed by Eve on BB84 protocol can give her $I_{EA}^{BB84}=I_{EB}^{BB84}=H(Q)$ \cite{Renner05}, where $H(Q)$ is Shannon entropy and $Q$ represents the level of QBER measured by Alice and Bob in their raw key. Since for the asymptotic case of infinitely long key, which we assume here, the mutual information between Alice and Bob when they are not performing preprocessing stage can be written as $I_{AB}=1-H(Q)$ \cite{Scarani09}, using equations (\ref{eq:keyDV}) and (\ref{eq:DeltaImain}) we can get the following expression for the lower bound on the secure key rate:
\begin{equation}
K^{(BB84)}=p_{exp}\max[0,1-2H(Q)].
\label{eq:DeltaIBB84}
\end{equation}

On the other hand, the upper bound on the information Eve can get by making the most general collective attacks when Alice and Bob use six-state protocol can be written as \cite{Renner05}
\begin{equation}
I_{EA}^{6state}=I_{EB}^{6state}=F(Q)-H(Q),
\end{equation}
where 
\begin{equation}
F(Q)=-\left(1-\frac{3Q}{2}\right)\log_2\left(1-\frac{3Q}{2}\right)-\frac{3Q}{2}\log_2\frac{Q}{2}
\end{equation}
If so, then from (\ref{eq:keyDV}) and (\ref{eq:DeltaImain}) we get 
\begin{equation}
 K^{(6state)}=p_{exp}\max\left[0,1-F(Q)\right]
\label{eq:DeltaI6state}.
\end{equation}

The above formulae for $K^{(BB84)}$ and $K^{(6state)}$ get more complicated, when Alice and Bob perform preprocessing, which can be done \emph{e.g.} by randomly flipping some bits of the raw key by Alice \cite{Kraus05,*Renner05}. In this case the mutual information about the key shared by Alice and Bob transforms into 
\begin{equation}
I_{AB}(Q,x)=1-H[(1-x)Q+x(1-Q)],
\end{equation}
where $x$ is the probability for Alice to flip a given bit of the raw key. In turn $I_{EA}^{BB84}$ (which is still equal to $I_{EB}^{BB84}$) can be written as
\begin{widetext}
\begin{eqnarray}
I_{EA}^{BB84}(Q,x)&=&\max_{x\in[0,1/2]}\min_{\lambda\in[0,Q]}\left[\sum_{i=1}^4A_i\log_2A_i-(1+\lambda-2Q)\log_2(1+\lambda-2Q)-\nonumber\right.\\&-&\left.2(Q-\lambda)\log_2(Q-\lambda)-\lambda\log_2\lambda\right],
\label{eq:IAEBB84preprocessing}
\end{eqnarray}
where 
\begin{equation}
A_{1,2}=\frac{1-Q\pm\sqrt{(1-Q)^2+16x(1-x)(\lambda-2Q+1)(\lambda-Q)}}{2}
\end{equation}
\end{widetext}
and
\begin{equation}
A_{3,4}=\frac{Q\pm\sqrt{Q^2+16x(1-x)\lambda(\lambda-Q)}}{2},
\end{equation}
while for six-state protocol we have
\begin{equation}
I_{EA}^{6state}(Q,x)=\max_{x\in[0,1/2]}\left[\sum_{i=1}^{4}B_i\log_2B_i+F(Q)\right],
\label{eq:IAE6statepreprocessing}
\end{equation}
where 
\begin{equation}
B_{1,2}=\frac{1-Q\pm\sqrt{(1-Q)^2-4x(1-x)Q(2-3Q)}}{2}
\end{equation}
and
\begin{equation}
B_{3,4}=\frac{Q\left[1\pm(1-2x)\right]}{2}.
\end{equation}

From the above analysis follows that the only parameter which Alice and Bob have to estimate in order to be able to assess the security of their DV QKD protocol is $Q$. We will further express this quantity in terms of the parameters of a given setup, taking into consideration the assumptions that were made at the beginning of this section. To do so let us first observe that since the scheme shown in Fig.\,\ref{fig:DVscheme} is perfectly symmetric in respect to polarizations, we don't have to consider separately the cases when Alice generates differently polarized photons. Instead of this, we can just consider one single case, in which Alice emits single photon in a randomly chosen polarization state, which we simply call \emph{right}. The orthogonal polarization state is called \emph{wrong} in this situation. Similarly, we call the detector to which signal photon emitted by Alice would go, if it is not lost during the propagation and if Bob chose the right basis for his measurement, the \emph{right} detector, and the other one -- the \emph{wrong} detector. Now by $p_+(k,l)$ [$p_-(k,l)$] let's denote the probability that signal photon would [would not] arrive at the right detector in a given attempt to generate a single bit of the key and at the same time $k$ noise photons would arrive at the right detector, while $l$ noise photons would arrive at the wrong detector. These two quantities are equal to
\begin{equation}
p_+(k,l)=T\pi_k(T)\pi_l(T)
\label{eq:pplus}
\end{equation}
and
\begin{equation}
p_-(k,l)=(1-T)\pi_k(T)\pi_l(T),
\label{eq:pminus}
\end{equation}
where
\begin{equation}
\pi_k(T)=\sum_{n=k}^\infty p_n(\mu){n \choose k} (1-T)^kT^{n-k}.
\end{equation}
Since we assume here that Bob's detectors do not have photon-number resolution, Alice and Bob automatically have to accept every situation in which all of the photons leaving the channel enter the same detector. Nevertheless, they can discard from the generated key all of the cases when both Bob's detectors click at the same time (we call this kind of event a \emph{double click} here). If they do so, the expected probability for accepting a given event by users of six-state protocol can be written as
\begin{equation}
p_{exp}=\sum_{k=0}^\infty p_+(k,0)+\sum_{k=1}^\infty p_-(k,0)+\sum_{l=1}^\infty p_-(0,l).
\label{eq:pexpII}
\end{equation}
It is clear that only the last term in the above formula contributes to the error rate, so the expression for QBER in our model takes the following form:
\begin{equation}
Q=\frac{\sum_{l=1}^\infty p_-(0,l)}{p_{exp}}.
\label{eq:QII}
\end{equation}

\section{Numerical results and analytical expressions}
\label{Sec:NumericalResults}

\begin{figure}[tbp]
\centering
\includegraphics[width=1.0\linewidth]{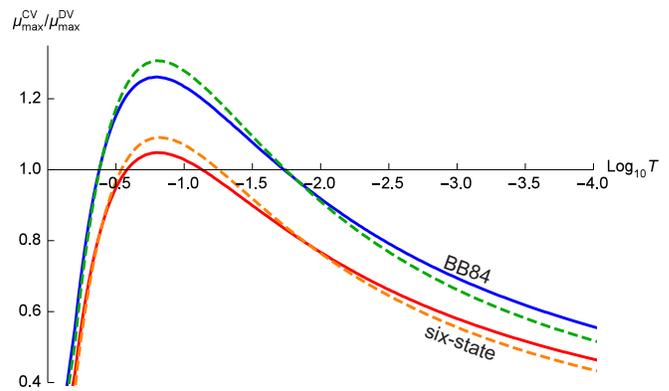}
\caption{(color online) Ratios between maximal values of $\mu$ for which it is possible to generate secure key using CV squeezed-state protocol ($\mu_\mathrm{max}^\mathrm{CV}$) and both DV  protocols ($\mu_\mathrm{max}^\mathrm{DV}$) considered in our analysis, plotted as a function of channel transmission $T$ for the situation when Alice and Bob perform the randomization stage of their raw key in order to increase its security (dashed lines) or do not perform it (solid lines).}
\label{fig:relativeresults}
\end{figure}

We now compare the security of the CV and DV QKD protocols in the presence of channel noise. To do so we perform numerical calculations in order to find the dependency of the maximal values of $\mu$, for which it is possible to generate secure key, on the transmittance $T$ of the channel connecting Alice and Bob in the cases when they use different QKD schemes. The relationships between such $\mu_\mathrm{max}^\mathrm{DV}(T)$ functions computed for BB84 and six-state protocols and the analogous function $\mu_\mathrm{max}^\mathrm{CV}(T)$ calculated for the squeezed-state scheme, both for the basic scenario and for the case when Alice and Bob try to improve the security of all these protocols by deliberately adding some noise to their raw keys (as was described in Sec.\,\ref{Sec:Models}), are presented in Fig.\,\ref{fig:relativeresults}.

Let us begin the analysis of Fig.\,\ref{fig:relativeresults} by focusing on the comparison between the six-state and squeezed-state protocols. As we can see in this picture for relatively high values of $T$ the former of these two cryptographic schemes allows for significantly higher values of $\mu$ than the latter one. However, this advantage quickly vanishes when $T$ decreases and for some intermediate values of the transmittance of the channel connecting Alice and Bob squeezed-state protocol appears to be slightly better suited for noisy quantum cryptography than the six-state scheme. Nevertheless, when $T$ decreases even further, at some point six-state protocol again starts to outperform the squeezed-state scheme and its advantage grows while $T\rightarrow 0$.

In fact, the relationship between BB84 and squeezed-state protocols is also very similar to the one described above. However, since for every value of $T$ BB84 protocol happens to be less resistant to the channel noise than the six-state scheme, the region of channel transmittance for which squeezed-state protocol allows for stronger channel noise than BB84 turns out to be significantly larger than in the case of the comparison between the six-state and squeezed-state protocols discussed before. Also the relative advantage of the CV protocol in this region is higher. In Fig.\,\ref{fig:relativeresults} we can also see that for the values of $T$ between roughly $10^{-0.5}$ and $10^{-2}$ adding noise to the raw key by the legitimate participants of a given QKD protocol can be more profitable for squeezed-state protocol than for DV protocols, while for $T<10^{-2}$ the situation is opposite.

Although it is not possible to find any simple, analytical expressions for the functions $\mu_{max}(T)$ in the general case, the analytical boundaries approximating it in the limit of $T\rightarrow 0$ can be derived for every protocol of our interest.

\textbf{Expression for DV QKD:} Derivation of the boundary for the case of six-state and BB84 protocols is relatively easy. To do it, we observe that when $T\rightarrow 0$ and $\mu\rightarrow 0$, the formula for QBER can be easily transformed into
\begin{equation}
Q\approx\frac{\mu}{2\mu+T}.
\end{equation}
If so, then for $T \ll 1$ the maximal secure value of $\mu$ depends on $T$ as follows:
\begin{equation}
\mu_{\mathrm max}(T)=\frac{TQ_\mathrm{th}}{1-2Q_\mathrm{th}},
\label{eq:approxmuDV}
\end{equation}
where $Q_\mathrm{th}$ is the threshold value of QBER, which for the cases of six-state and BB84 protocols are approximately equal to $12.6\%$ and $11\%$ respectively \cite{Renner05}.

\textbf{Expression for CV QKD:} In the case of the squeezed-state CV QKD protocol, when no noise is deliberately added on the receiver side, the analytical lower bound on the secure key rate can be simplified to
\begin{equation}
K^{(CV)} \approx (T-\mu)\log_2{e}+\mu\log_2{\mu},
\end{equation}
by using series expansion around $T=0$, taking the limit of infinite modulation $V \to \infty$ and performing series expansion around $\mu=0$. The value of $\mu$ which turns this simplified expression to zero can be calculated analytically and expressed using Lambert W function as
\begin{equation}
\mu_{\mathrm max}(T)=\exp[1+W_{-1}(-T/e)].
\label{eq:approxmuCV}
\end{equation}

\begin{figure}[tpb]
\centering
\includegraphics[width=1.0\linewidth]{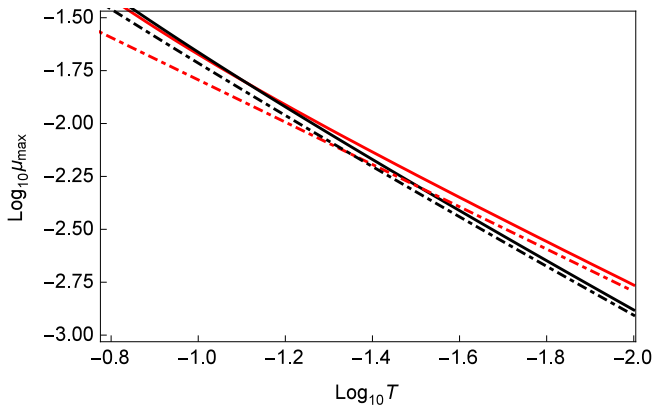}
\caption{(color online) Maximal values of $\mu$ for which it is possible to generate secure key as a function of channel transmittance $T$ calculated numerically (solid lines) for the cases of Alice and Bob using six-state protocol (red lines) and squeezed-state protocol (black lines), plotted along with the analytical approximations (dot-dashed lines) of the functions $\mu_{\mathrm max}(T)$ valid for the case of $T\rightarrow0$, given by formulae (\ref{eq:approxmuDV}) and (\ref{eq:approxmuCV}) respectively.}
\label{fig:zoomresults}
\end{figure}

The comparison between the boundaries given by formulae (\ref{eq:approxmuDV}) and (\ref{eq:approxmuCV}) and the results of our numerical calculations of the functions $\mu_{\mathrm max}(T)$ performed for the cases of six-state and squeezed-state protocols, which is illustrated in Fig.\,\ref{fig:zoomresults}, shows good agreement between our analytical and numerical results in the limit of $T\rightarrow0$.

\begin{figure}
\centering
\includegraphics[width=1.0\linewidth]{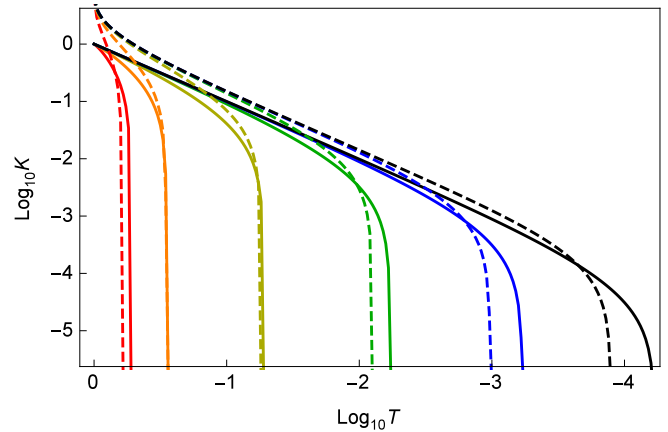}
\caption{(color online) Lower bound for the secure key rate as a function of transmittance of the channel connecting Alice and Bob, plotted for $\mu=0.5$ (red lines), $\mu=10^{-1}$ (orange lines), $\mu=10^{-2}$ (yellow lines), $\mu=10^{-3}$ (green lines), $\mu=10^{-4}$ (blue lines) and $\mu=10^{-5}$ (black lines) for six-state protocol (solid lines) and squeezed-state protocol (dashed lines) with the assumption that Alice's sources and Bob's detection systems are perfect.}
\label{fig:KeyComparison}
\end{figure}

While our main goal in performing the analysis presented above was to identify the conditions in which only one of the two main families of QKD protocols can be used to provide security for the process of key generation, its results cannot help us with answering the question which protocol one should choose in a particular case when both CV and DV QKD schemes can be secure at the same time. Facing such a decision it is always good to compare the lower bounds for the secure key rate for different protocols. This kind of comparison, performed for the six-state and squeezed-state schemes, is presented in Fig.\,\ref{fig:KeyComparison} where the function of $K(T)$ was plotted for a few different values of $\mu$, ranging from $10^{-5}$ to $0.5$. Although typical values of $\mu$ in a dark fiber, dedicated solely for the generation of secret key, can be estimated to be on the level of $10^{-4}$--$10^{-5}$ (basing on the experimental results obtained in \cite{Jouguet2013}), in commercial QKD applications utilizing telecom fibers populated by strong classical signals the channel noise can be considerably stronger. In this situation the actual level of $\mu$ would primarily depend on the number of classical channels multiplexed in a given fiber and the power of classical signals transmitted through them \cite{Qi10}. For this reason in the analysis presented in this paper we decided not to focus on a particular level of $\mu$, but consider a broad range of its values, encompassing several orders of magnitude.

From the Fig.\,\ref{fig:KeyComparison} one can conclude that if only $T$ is considerably larger than the minimal secure transmittance  of the channel connecting Alice and Bob for CV squeezed-state protocol, this scheme can always provide comparable but slightly higher lower bound on the secure key rate than the six-state DV QKD protocol. Similar conclusion can be drawn from the comparison of BB84 and squeezed-state schemes. The main reason for this advantage stems from the capability of encoding more than one bit of information in a single pulse by using CV QKD protocols, which in turn is impossible for the considered DV schemes based on qubits. The results presented in Fig.\,\ref{fig:KeyComparison} can be also used to predict the outcome of a possible comparison of the robustness of the six-state and squeezed-state protocols to the channel noise for a given non-zero value of $K$. In this case one should just compare the minimal secure values of $T$ for these protocols, which can be reached for different levels of $\mu$ for a desired $K$. It is important to note, however, that the lower bound on the secure key rate in our work is calculated per use of the channel, so it contains only partial information on the achievable rate of a particular implementation of a given QKD protocol. In order to calculate the lower bound on the amount of bits of the final key per unit of time, one would have to multiply the expression for $K$ (formula (\ref{eq:keyDV}) or (\ref{eq:keyCV}) for the DV or CV protocols respectively) by the repetition rate of the system, which depends on the setup. Therefore comparing the key rates in the general case can be misleading.

\begin{figure}
\centering
\includegraphics[width=1.0\linewidth]{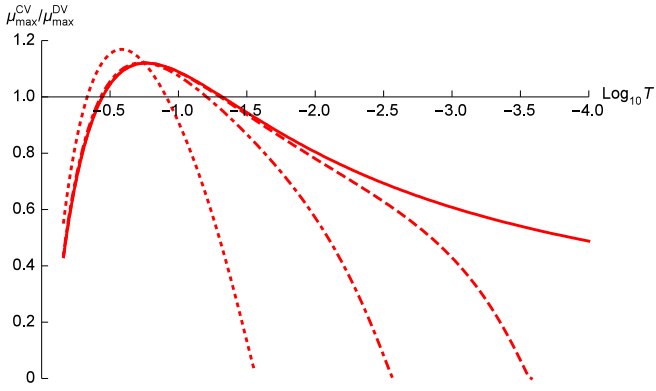}
\caption{(color online) Ratio between maximal values of $\mu$ for which it is possible to generate secure key using squeezed-state protocol ($\mu_\mathrm{max}^\mathrm{CV}$) and six-state protocol ($\mu_\mathrm{max}^\mathrm{DV}$), plotted as a function of channel transmittance $T$ for the asymptotic case in which the number of quantum signals exchanged by Alice and Bob during the protocol is infinite (solid line) and for the situations when it equals to $N=10^{10}$ (dashed line), $N=10^8$ (dot-dashed line) and $N=10^6$ (dotted line). The calculations were made with the assumption that the trusted parties do not increase the security of their raw key by performing the randomization stage.}
\label{fig:MuComparisonFinite}
\end{figure}

The analysis presented above was performed for the asymptotic case of infinite number of quantum signals exchanged by Alice and Bob during the key generation process. However, in realistic situation this number, denoted here by $N$, is always finite. Therefore, it is instructive to check the stability of the discussed results in the finite-key regime. In order to do that we utilize the calculation method introduced for the DV QKD case in \cite{Scarani08} and adopted for the CV protocols in \cite{Leverrier2010}. For definiteness we set the values of all of the failure probabilities present in the mathematical formulas introduced there to the level of $10^{-10}$. The results of this calculation are illustrated in Fig.\,\ref{fig:MuComparisonFinite}, where the ratio of $\mu^\mathrm{CV}_\mathrm{max}/\mu^\mathrm{DV}_\mathrm{max}$ for squeezed-state and six-state protocols is plotted for different numbers of $N$. 

As it turns out, if only the transmittance of the quantum channel connecting Alice and Bob is not particularly high, the finite-size effects have more negative influence on the squeezed-state protocol than on the DV schemes. In particular, for any finite $N$ there exists a corresponding threshold value for $T$ below which generation of secure key by utilizing squeezed-state protocol becomes impossible even for $\mu\rightarrow 0$. On the other hand, as long as the lossy and noisy channel connecting Alice and Bob is the only imperfect setup element, no such threshold appears for the DV protocols. Therefore, for limited $N$ the ratio of $\mu^\mathrm{CV}_\mathrm{max}/\mu^\mathrm{DV}_\mathrm{max}$ decreases much faster and eventually reaches zero, contrary to the asymptotic case. Furthermore, Fig.\,\ref{fig:MuComparisonFinite} shows that the value of $T$, below which six-state protocol becomes more resistant to the channel noise than a given CV QKD scheme, grows with the decreasing number of quantum signals exchanged by Alice and Bob.

\section{Requirements for nonclassical resources}
\label{Sec:RealisticCase}

Knowing that for the case when Alice's source and Bob's detection system are perfect DV QKD protocols can provide one with the security of key generation process for slightly higher values of $\mu$ than the squeezed-state CV protocol if only the transmittance of the channel connecting Alice and Bob is low enough, we can now consider the possibility for realizing this kind of scenario in the situation when the sources of signal owned by Alice are not ideal. 

In order to assess the quality of a single-photon source needed for secure realization of QKD protocols for the combinations of parameters $T$ and $\mu$ for which squeezed-state protocol is insecure, we will assume in this section that Alice's source produces genuine single-photon pulses with probability $p$ and empty pulses with probability $1-p$, \emph{i.e.}\,it never emits multiphoton pulses. The reason for adopting this particular model of Alice's source for our considerations is that while decreasing the probability for multiphoton emission to a very low level is possible these days for many different kinds of realistic single-photon sources \cite{Fasel04, Keller04,*Brokmann04,*Laurat06,*Pisanello10,*Mucke13,Claudon10}, constructing a high-quality source which would produce non-empty pulses with probability close to one remains a serious challenge for experimental physicists. This task is  especially hard to be accomplished for the case of deterministic single-photon sources, which are usually affected by poor collection efficiency of generated photons \cite{Eisaman11}. However, very promising sources based on quantum dots embedded in photonic nanowires or micropillar cavities have been developed recently, with probability of producing a single-photon pulse exceeding $70\%$ and potentially reaching even $95\%$ \cite{Claudon10,Bulgarini12,*Gazzano12,Somaschi16}. Furthermore, relatively efficient probabilistic single-photon sources, based especially on the spontaneous parametric down-conversion (SPDC) process, with very low probabilities of emitting a multiphoton pulse and with the heralding efficiency exceeding $60\%$ were already developed more than a decade ago \cite{Fasel04}. Nowadays, reports on SPDC-based sources with $p>80\%$ can be find in the literature \cite{Pomarico12,*Pereira13,*Ramelow13}.

\begin{figure}
\centering
\includegraphics[width=1\linewidth]{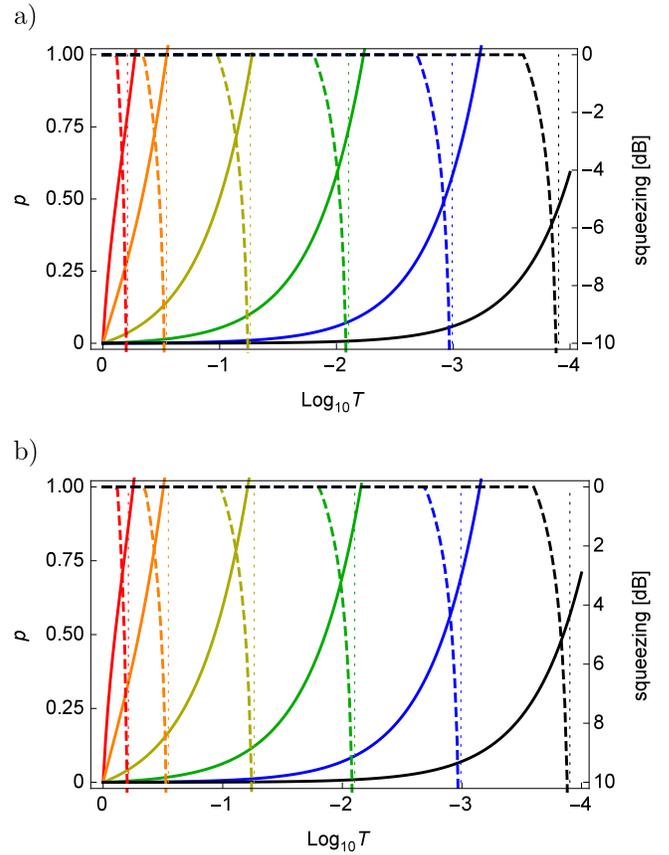}
\caption{(color online) Requirements on the value of squeezing parameter (dashed lines) and the probability $p$ of producing non-empty signal pulse by a single-photon source (solid lines) needed to be reached for the security of, respectively, the squeezed-state protocol and a) six-state, b) BB84 protocol, plotted as functions of the transmittance of the channel connecting Alice and Bob for six different values of $\mu$: $\mu=0.5$ (red lines), $\mu=10^{-1}$ (orange lines), $\mu=10^{-2}$ (yellow lines), $\mu=10^{-3}$ (green lines), $\mu=10^{-4}$ (blue lines) and $\mu=10^{-5}$ (black lines). Vertical dotted lines denote the values of $T$ for which squeezed-state protocol becomes insecure for particular values of $\mu$.}
\label{fig:CVDVrealistic}
\end{figure}

\begin{figure}
\centering
\includegraphics[width=1\linewidth]{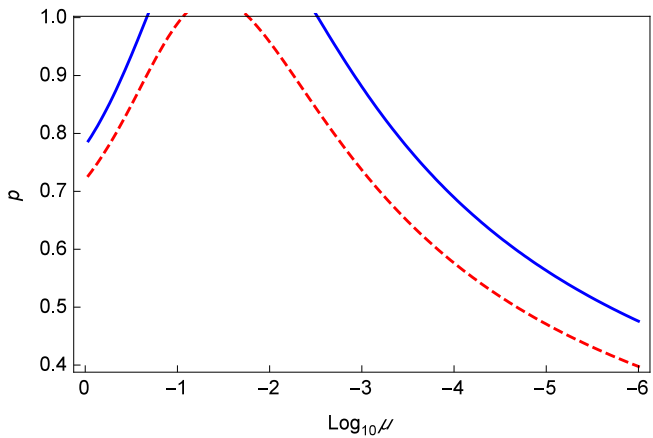}
\caption{(color online) Requirements on the probability $p$ of producing non-empty signal pulse by a single-photon source needed to be fulfilled for the generation of secure key by using six-state (red, dashed line) or BB84 (blue, solid line) protocol for the same value of $\mu$, for which squeezed-state protocol with infinite squeezing stops being secure at a given transmittance $T$.}
\label{fig:DVReqComparison}
\end{figure}

Adopting the model for realistic single-photon source described above, we investigated the dependence of the minimal probability $p$ of producing non-empty pulse by Alice's source, required for the six-state and BB84 protocols to be secure, on the transmittance of the channel connecting Alice and Bob for a few different values of the power of the source of noise in the DV QKD model, illustrated in Fig.\,\ref{fig:DVscheme}.
The results of this investigation are plotted in Fig.\,\ref{fig:CVDVrealistic}. In the same figure we also plotted the dependency of the value of squeezing parameter, required for the security of the squeezed-state protocol in the model for CV QKD pictured in Fig.\,\ref{fig:CVscheme}, on $T$. In order to make necessary calculations for squeezed-state protocol in realistic case we used the generalized state preparation model for CV QKD in which modulation and squeezing of the states emitted by Alice's source can be parametrized separately \cite{Usenko2011}. While the plots given in Fig.\,\ref{fig:CVDVrealistic} were obtained for the very strong modulation variance ($10^3$ shot-noise units), varying this quantity does not significantly affect the results.

From Fig.\,\ref{fig:CVDVrealistic} one can deduce that the requirements for the probability $p$ of emitting non-empty pulse by Alice's single-photon source, which would have to be fulfilled in order to ensure security of the DV QKD protocols for the values of $T$ for which squeezed-state protocol is no longer secure, are generally quite demanding, especially if the power of the source of noise is relatively high. For different levels of $\mu$ the minimal values of $p$ which would be needed to realize this task are given by the crossing points of the solid and dotted lines of the same colors displayed in Fig.\,\ref{fig:CVDVrealistic}. While in practice overcoming the squeezed-state protocol by the DV QKD schemes may be very hard or even impossible to demonstrate for relatively high values of $\mu$, it is certainly achievable for realistic sources in the case of $\mu\ll 1$, as the requirements for $p$ shown in Fig.\,\ref{fig:CVDVrealistic} become more and more relaxed when $\mu\rightarrow 0$. This conclusion can be confirmed in Fig.\,\ref{fig:DVReqComparison}, where the  minimal required values of $p$ are plotted as the functions of $\mu$ both for the BB84 and six-state protocol. The results of our analysis shown in Fig.\,\ref{fig:CVDVrealistic} and Fig.\,\ref{fig:DVReqComparison} indicate that even DV QKD schemes with inefficienct sources of photons can be capable to overcome the CV protocols for long-distance quantum cryptography with ultra low channel noise.

Not surprisingly, in Fig.\,\ref{fig:DVReqComparison} one can also see that for every level of $\mu$ the value of $p$ needed to overcome squeezed-state protocol is larger for the BB84 than for the six-state protocol. This means that a demonstration of the superiority of the six-state protocol over the squeezed-state scheme in realistic situation would be easier to perform than an analogous demonstration for BB84 protocol. This conclusion justifies our choice to focus more on the six-state protocol in this work, despite much larger popularity of the BB84 scheme.

\begin{figure}
\centering
\includegraphics[width=1\linewidth]{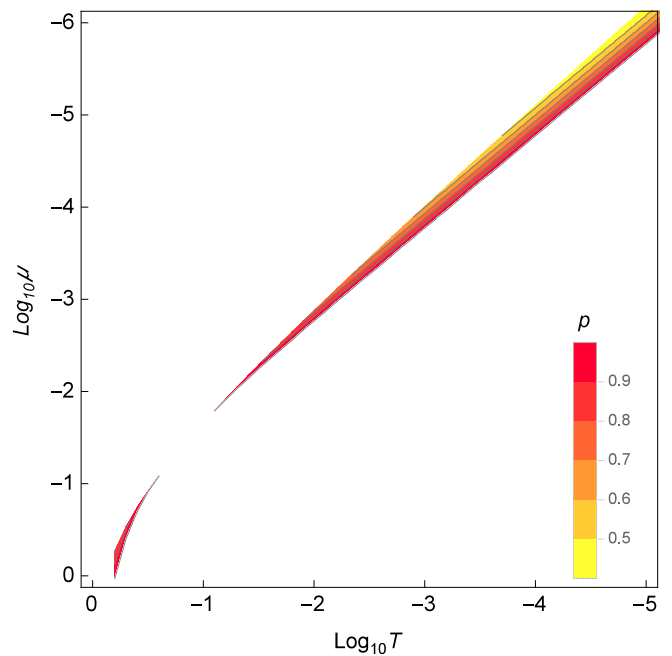}
\caption{(color online) Minimal values of the probability $p$ of producing non-empty signal pulse by a single-photon source, needed for the six-state protocol to be secure for a given pair of values of the channel transmittance $T$ and noise mean photon number $\mu$ for which squeezed-state protocol is already insecure. White color indicates the regions of the plot where either the squeezed-state protocol is still secure or the six-state protocol is insecure even for $p=1$.}
\label{fig:p1ContourPlot}
\end{figure}

While Fig.\,\ref{fig:DVReqComparison} shows only the minimal values of $p$ for which DV QKD protocols can still be secure for given $\mu$ and $T$ that already breaks the security of the CV QKD schemes, for higher $p$ demonstrating the superiority of BB84 or six-state protocol over the squeezed-state scheme may be realized also for the lower transmittance of the quantum channel connecting Alice and Bob. Therefore, it is reasonable to ask about the whole region of parameters $\mu$ and $T$ for which overcoming the performance of the squeezed-state protocol by a given DV QKD scheme is possible. Such a region, found for the case of six-state protocol, is illustrated in Fig.\,\ref{fig:p1ContourPlot}. One can see there that it is relatively narrow. This is because the closer $T$ is to the minimal secure transmittance of the quantum channel connecting Alice and Bob for a given $\mu$, the faster the minimal required value of $p$ goes to one. This tendency could actually be observed even before, in Fig.\,\ref{fig:CVDVrealistic}. Fig.\,\ref{fig:p1ContourPlot} also confirms that the requirement for $p$ relaxes when $\mu\rightarrow0$. 

Beside sources of photons, another fundamental part of the setup needed for the implementation of DV QKD protocols are single-photon detectors. In some situations imperfection of these devices can also affect the security of such schemes in significant way. In particular, every realistic single-photon detector is characterized by a non-zero dark count rate. The influence of these unwanted clicks on the results presented in this work is negligible as long as the value of $T$ is more than two orders of magnitude higher than the probability $d$ to register a dark count per single detection window.  However, for lower transmittance of the quantum channel dark counts considerably affect the security of DV QKD protocols and can become the major issue. They result in threshold values of channel transmittance $T_{th}$, below which overcoming squeezed-state protocol with DV schemes becomes impossible even if the single-photon source used by Alice is perfect. These thresholds strongly depend  on the relationship between $d$ and the detection efficiency $\eta$ of the measurement devices utilized by Bob. Typical values of $d/\eta$ that can be found in the literature describing recent DV QKD experiments range from $10^{-4}$ to $10^{-7}$ \cite{Zhang08,*Walenta14,*Valivarthi15,*Takemoto15,*Wang15,*Tang16}. During our work we found out that in this region $T_{th}$ can be upper-bounded by 
\begin{equation}
T_{th}^{\,6 state}\leq10^{1.07\log_{10}\left(d/\eta\right)+1.45}
\end{equation} 
for the case when the trusted parties implement six-state protocol or 
\begin{equation}
T_{th}^{BB84}\leq10^{1.15\log_{10}\left(d/\eta\right)+2.12}.
\end{equation}
when they choose BB84 scheme.

On the other hand, if Bob's measurement system does not register any dark counts, the limited detection efficiency does not affect the results of our calculations as long as $T<10^{-2}$. This is because for low values of $T$ almost all of the non-empty pulses arriving at Bob's measurement system contain either a single signal photon or a single noise photon. Therefore, since the limited detection efficiency reduces the fractions of registered signal and noise photons in exactly the same way its value does not matter for the security threshold. Only when the transmittance of the quantum channel connecting Alice and Bob is relatively high and the probability for more photons to arrive at Bob's detectors at the same time becomes significant, the situation can be different. In this case limited detection efficiency makes the requirement for the quality of Alice's source slightly more demanding.

\section{Conclusions}
\label{Sec:Conclusions}

In the analysis presented above we compared the security of two DV protocols, namely BB84 and six-state, and CV squeezed-state protocol in the situation when the only imperfect element of the setup used by Alice and Bob is the quantum channel connecting them. We assumed here that this channel is lossy and that the noise coupled to the signal during its propagation through it is of the type of thermal reservoir, which can be seen as a typical scenario for CV QKD case. The results of our analysis, depicted in Fig.\,\ref{fig:relativeresults}, clearly show that while for some intermediate values of the channel transmittance continuous-variable squeezed-state protocol is comparably resilient to the channel noise as BB84 and six-state schemes, for the cases of $T \to 1$ and $T \ll 1$ both the DV protocols perform better. It suggests that in the scenario when Alice and Bob have high-quality sources and detectors, but the quantum channel connecting them is lossy and noisy, DV QKD technique can be seen as having more potential for generating a secure cryptographic key than CV QKD. Although exploiting this potential in practice may be challenging, it is within our reach. With the recent engineering progress in the field of single-photon sources it can be even possible to demonstrate the superiority of realistic DV protocols over the infinite-squeezing ideal CV schemes in the regime of $T \ll 1$, as can be seen in Fig.\,\ref{fig:CVDVrealistic}. This conclusion may provide some additional motivation for the experimental physicists to focus even more of their efforts on developing novel high-quality sources with high probability of producing non-empty pulse and very low probability for multiphoton emission or improving the performance of the existing ones.

\medskip
\noindent {\bf Acknowledgments. --} The research leading to these results has received funding from the EU FP7 under Grant Agreement No. 308803 (project BRISQ2), co-financed by M\v SMT \v CR (7E13032). M.L. and V.C.U. acknowledge the project 13-27533J of the Czech Science Foundation. M.L. acknowledges support by the Development Project of Faculty of
Science, Palacky University.

\appendix

\section{}
\label{Sec:GG02protocol}

\begin{figure}[tbp]
\centering
\includegraphics[width=1.0\linewidth]{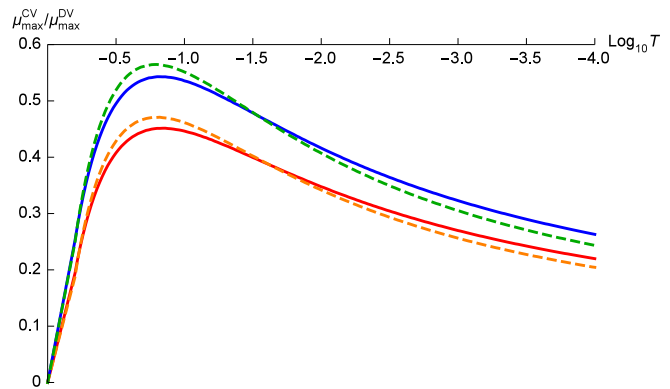}
\caption{(color online) Ratios between maximal values of $\mu$ for which it is possible to generate secure key using CV GG02 protocol ($\mu_\mathrm{max}^\mathrm{CV}$) and both DV  protocols ($\mu_\mathrm{max}^\mathrm{DV}$) considered in our analysis, plotted as a function of channel transmission $T$ for the situation when Alice and Bob perform the randomization stage of their raw key in order to increase its security (dashed lines) or do not perform it (solid lines).}
\label{fig:relativeresults2}
\end{figure}

In the main body of our paper we considered Gaussian squeezed-state CV QKD protocol, using it for the comparison with DV QKD protocols in terms of their robustness to the channel noise. However, due to the popularity of the GG02 scheme based on coherent states \cite{Grosshans2002}, it is meaningful to perform a similar analysis also for this protocol. In order to do all the necessary calculations in this situation, one can once again utilize the formulae introduced in Sec.\,\ref{Sec:CVmodel}, only assuming  that this time the variance of the signal states is $V=1$. In Fig.\,\ref{fig:relativeresults2} we present the results of our comparison between the maximal values of the parameter $\mu$ ensuring the security of GG02 and the DV QKD protocols. This comparison is similar to the one made for the squeezed-state scheme in Sec.\,\ref{Sec:NumericalResults}, which results are depicted in Fig.\,\ref{fig:relativeresults}. By comparing the two aforementioned figures with each other one can confirm that the squeezed-state protocol is indeed more resistant to the channel noise than the GG02 scheme, as was already stated in the first paragraph of Sec.\,\ref{Sec:CVmodel}. As can be seen in Fig.\ref{fig:relativeresults2}, contrary to the case of the squeezed-state protocol, for every possible value of $T$, GG02 scheme allows for significantly lower values of $\mu_\mathrm{max}$ than the BB84 and six-state protocols.

\section{}
\label{Sec:DifferentDetection}

In the analysis of DV QKD protocols presented in the main body of this article we assumed that Bob's detectors have perfect detection efficiency, but do not have the ability to resolve the number of photons entering them. At first sight it would seem that replacing them with photon-number-resolving detectors should improve the setup, making it more resilient to the channel noise. However this intuition does not necessarily has to be correct. Here we are going to show that in our model, when the source of channel noise has thermal statistics, equipping Bob's detectors with photon-number resolution does not change the function of $\mu_\mathrm{max}^\mathrm{DV}(T)$ in any way, while for Poisson statistics it can even have negative effect on QKD security.

In order to accomplish this task, we will start with adapting the expressions for $p_{exp}$ and $Q$, given previously by formulae (\ref{eq:pexpII}) and (\ref{eq:QII}) respectively, to the case of photon-number-resolving detetcors used by Bob. We get:
\begin{equation}
p_{exp}^{(II)}= p_+(0,0)+p_-(1,0)+p_-(0,1)
\end{equation}
and
\begin{equation}
Q^{(II)}=\frac{ p_-(0,1)}{p_{exp}^{(II)}}.
\label{eq:QIII}
\end{equation}
Since for both DV protocols considered here the formulae for $\Delta I$ depend only on the parameter $Q$ (and optinally the probability $x$ to flip a bit by Alice, if the preprocessing stage is being performed), it is obvious that the condition for photon-number-resolving detectors to offer better security of our DV QKD schemes than simple on/off binary detectors can be written in the form of the following inequality:
\begin{equation}
Q^{(II)}<Q.
\end{equation}
Using equations (\ref{eq:QII}) and (\ref{eq:QIII}), and taking advantage of the facts that
\begin{equation}
p_+(k,l)=p_+(l,k)
\end{equation}
and
\begin{equation}
p_-(k,l)=\frac{1-T}{T}p_+(k,l),
\end{equation}
we can transfrom this condition into
\begin{equation}
p_+(1,0)\cdot\sum_{k=1}^\infty p_+(k,0)<p_+(0,0)\cdot\sum_{k=2}^\infty p_+(k,0).
\end{equation}
After inserting (\ref{eq:pplus}) and performing some algebraic calculations, we can get the following final version of this condition:
\begin{eqnarray}
&&\sum_{k=1}^\infty\sum_{n=k}^\infty\sum_{m=1}^\infty\left[p_n(\mu)p_m(\mu)m{n \choose k}-\right.\\&-&\left.p_{n+1}(\mu)p_{m-1}(\mu){n+1 \choose k+1}\right](1-T)^kT^{n+m-k-1}<0\nonumber.
\label{eq:condition2}
\end{eqnarray}

The above inequality cannot be solved analytically in the general case. However, it can be further simplified in two extreme cases of $T\rightarrow 0$ and $T\rightarrow 1$. If $T\rightarrow 0$, we can leave only the expression for $m=1$ and $n=k$ on the left-hand side of the condition (\ref{eq:condition2}). If we do it, we get:
\begin{equation}
\sum_{k=1}^\infty\left[p_k(\mu)p_1(\mu)-p_{k+1}(\mu)p_{0}(\mu)\right]<0,
\label{eq:condition3}
\end{equation}
But for the thermal statistics we have
\begin{equation}
p_k(\mu)p_1(\mu)-p_{k+1}(\mu)p_{0}(\mu)=0
\end{equation}
for every $k$. This means that equipping Bob's detectors with the ability to resolve the number of incoming photons does not have any effect on the function $\mu_\mathrm{max}(T)$ when $T\rightarrow 0$.

\begin{figure}[tbp]
\centering
\includegraphics[width=1.0\linewidth]{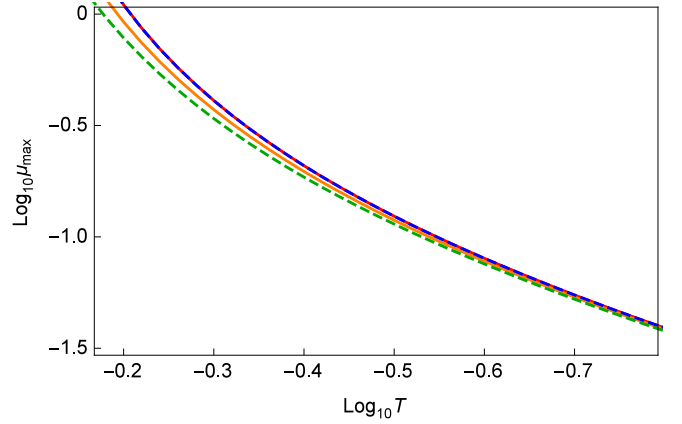}
\caption{(color online) Maximal values of $\mu$ for which it is possible to generate secure key as a function of channel transmission $T$ plotted for the case of Alice and Bob using six-state protocol when the channel noise has thermal statistics and the detectors used by Bob have photon-number-resolving ability (dashed blue line) or do not have it (solid red line). Analogous results for the Poissonian type of noise are plotted with dashed green line (for detectors with photon number resolution) and solid orange line (for detectors without photon number resolution).}
\label{fig:detectionmodels}
\end{figure}

The situation for $T\rightarrow 1$ is more complicated. In this case we can leave on the left-hand side of inequality (\ref{eq:condition2}) only the expression with lowest possible power of $(1-T)$, that is for $k=1$. Then we have
\begin{equation}
\sum_{n,m=1}^\infty\left[nmp_n(\mu)p_m(\mu)-{n+1 \choose 2}p_{n+1}(\mu)p_{m-1}(\mu)\right]<0
\label{eq:condition4}.
\end{equation}
For thermal statistics of the source of noise this condition becomes
\begin{equation}
\sum_{n,m=1}^\infty \frac{\mu^{n+m}}{(\mu+1)^{n+m+2}}n\left[m-\frac{n+1}{2}\right]<0.
\label{eq:inequality5}
\end{equation}
A good method to prove that the left hand side of this inequality is equal to zero is to show that for any $c$ the term standing beside $\mu^c/(\mu+1)^{(c+2)}$, which can be actually written as
\begin{equation}
\sum_{n=1}^{c-1}n\left[c-n-\frac{n+1}{2}\right],
\end{equation}
is equal to zero. This can be done by induction.

On the other hand for Poisson statistics of the sources of noise, inequality (\ref{eq:condition2}) would transform into
\begin{equation}
\sum_{k=1}^\infty e^{-2\mu}\mu^{k+1}\left[\frac{1}{k!}-\frac{1}{(k+1)!}\right]<0
\end{equation}
for the case of $T\rightarrow 0$ or into
\begin{equation}
\sum_{n,m=1}^\infty e^{-2\mu}\frac{\mu^{n+m}}{2(n-1)!(m-1)!}<0
\end{equation}
for the case of $T\rightarrow 1$. It is not difficult to see, that the left-hand sides of both these inequalities are actually larger than zero, which means that if the sources of noise in our DV QKD scheme had Poisson statistics, from the point of its resilience to noise it would be better for Bob to use simple on/off detectors instead of photon-number-resolving ones.

The conclusions which can be drawn from the above analysis can be confirmed in Fig.\,\ref{fig:detectionmodels}, where we present the results of the numerical calculations of the function $\mu_\mathrm{max}^\mathrm{DV}(T)$ for the cases of thermal and Poisson statistics of the source of noise both in the situation when Bob uses detectors with and without the ability to resolve the number of photons entering them.

In fact, it is quite easy to intuitively explain why using photon-number-resolving detectors by Bob does not seem to improve the security of our DV QKD scheme over the case of on/off detectors. The basic reason for this is that detectors with  photon-number resolution exclude from the key not only all the situations in which more than one photon comes to the wrong detector (which is obviously good for the security), but also all the cases when more than one photon arrives in the right detector (which is obviously bad). So although using photon-number-resolving detectors reduces the number of errors in the key, QBER given by the formula (\ref{eq:QII}) can actually increase due to even greater reduction of $p_{exp}$ at the same time.

\bibliography{refs4}

\begin{thebibliography}{98}%
\makeatletter
\providecommand \@ifxundefined [1]{%
 \@ifx{#1\undefined}
}%
\providecommand \@ifnum [1]{%
 \ifnum #1\expandafter \@firstoftwo
 \else \expandafter \@secondoftwo
 \fi
}%
\providecommand \@ifx [1]{%
 \ifx #1\expandafter \@firstoftwo
 \else \expandafter \@secondoftwo
 \fi
}%
\providecommand \natexlab [1]{#1}%
\providecommand \enquote  [1]{``#1''}%
\providecommand \bibnamefont  [1]{#1}%
\providecommand \bibfnamefont [1]{#1}%
\providecommand \citenamefont [1]{#1}%
\providecommand \href@noop [0]{\@secondoftwo}%
\providecommand \href [0]{\begingroup \@sanitize@url \@href}%
\providecommand \@href[1]{\@@startlink{#1}\@@href}%
\providecommand \@@href[1]{\endgroup#1\@@endlink}%
\providecommand \@sanitize@url [0]{\catcode `\\12\catcode `\$12\catcode
  `\&12\catcode `\#12\catcode `\^12\catcode `\_12\catcode `\%12\relax}%
\providecommand \@@startlink[1]{}%
\providecommand \@@endlink[0]{}%
\providecommand \url  [0]{\begingroup\@sanitize@url \@url }%
\providecommand \@url [1]{\endgroup\@href {#1}{\urlprefix }}%
\providecommand \urlprefix  [0]{URL }%
\providecommand \Eprint [0]{\href }%
\providecommand \doibase [0]{http://dx.doi.org/}%
\providecommand \selectlanguage [0]{\@gobble}%
\providecommand \bibinfo  [0]{\@secondoftwo}%
\providecommand \bibfield  [0]{\@secondoftwo}%
\providecommand \translation [1]{[#1]}%
\providecommand \BibitemOpen [0]{}%
\providecommand \bibitemStop [0]{}%
\providecommand \bibitemNoStop [0]{.\EOS\space}%
\providecommand \EOS [0]{\spacefactor3000\relax}%
\providecommand \BibitemShut  [1]{\csname bibitem#1\endcsname}%
\let\auto@bib@innerbib\@empty
\bibitem [{\citenamefont {Bennett}\ and\ \citenamefont
  {Brassard}(1984)}]{Bennett84}%
  \BibitemOpen
  \bibfield  {author} {\bibinfo {author} {\bibfnamefont {C.~H.}\ \bibnamefont
  {Bennett}}\ and\ \bibinfo {author} {\bibfnamefont {G.}~\bibnamefont
  {Brassard}},\ }in\ \href@noop {} {\emph {\bibinfo {booktitle} {Proceedings of
  the IEEE International Conference on Computers, Systems, and Signal
  Processing, Bangalore, India}}},\ Vol.~\bibinfo {volume} {11}\ (\bibinfo
  {publisher} {IEEE, New York},\ \bibinfo {year} {1984})\ pp.\ \bibinfo {pages}
  {175--179}\BibitemShut {NoStop}%
\bibitem [{\citenamefont {Ekert}(1991)}]{Ekert1991}%
  \BibitemOpen
  \bibfield  {author} {\bibinfo {author} {\bibfnamefont {A.~K.}\ \bibnamefont
  {Ekert}},\ }\href@noop {} {\bibfield  {journal} {\bibinfo  {journal} {Phys.
  Rev. Lett.}\ }\textbf {\bibinfo {volume} {67}},\ \bibinfo {pages} {661}
  (\bibinfo {year} {1991})}\BibitemShut {NoStop}%
\bibitem [{\citenamefont {Ralph}(1999)}]{Ralph99}%
  \BibitemOpen
  \bibfield  {author} {\bibinfo {author} {\bibfnamefont {T.~C.}\ \bibnamefont
  {Ralph}},\ }\href@noop {} {\bibfield  {journal} {\bibinfo  {journal} {Phys.
  Rev. A}\ }\textbf {\bibinfo {volume} {61}},\ \bibinfo {pages} {010303}
  (\bibinfo {year} {1999})}\BibitemShut {NoStop}%
\bibitem [{\citenamefont {Weedbrook}\ \emph {et~al.}(2012)\citenamefont
  {Weedbrook}, \citenamefont {Pirandola}, \citenamefont
  {Garc\'{i}a-Patr\'{o}n}, \citenamefont {Cerf}, \citenamefont {Ralph},
  \citenamefont {Shapiro},\ and\ \citenamefont {Lloyd}}]{Weedbrook12}%
  \BibitemOpen
  \bibfield  {author} {\bibinfo {author} {\bibfnamefont {C.}~\bibnamefont
  {Weedbrook}}, \bibinfo {author} {\bibfnamefont {S.}~\bibnamefont
  {Pirandola}}, \bibinfo {author} {\bibfnamefont {R.}~\bibnamefont
  {Garc\'{i}a-Patr\'{o}n}}, \bibinfo {author} {\bibfnamefont {N.~J.}\
  \bibnamefont {Cerf}}, \bibinfo {author} {\bibfnamefont {T.~C.}\ \bibnamefont
  {Ralph}}, \bibinfo {author} {\bibfnamefont {J.~H.}\ \bibnamefont {Shapiro}},
  \ and\ \bibinfo {author} {\bibfnamefont {S.}~\bibnamefont {Lloyd}},\
  }\href@noop {} {\bibfield  {journal} {\bibinfo  {journal} {Rev. Mod. Phys.}\
  }\textbf {\bibinfo {volume} {84}},\ \bibinfo {pages} {621} (\bibinfo {year}
  {2012})}\BibitemShut {NoStop}%
\bibitem [{\citenamefont {Cerf}\ \emph {et~al.}(2001)\citenamefont {Cerf},
  \citenamefont {Levy},\ and\ \citenamefont {Van~Assche}}]{Cerf2001}%
  \BibitemOpen
  \bibfield  {author} {\bibinfo {author} {\bibfnamefont {N.~J.}\ \bibnamefont
  {Cerf}}, \bibinfo {author} {\bibfnamefont {M.}~\bibnamefont {Levy}}, \ and\
  \bibinfo {author} {\bibfnamefont {G.}~\bibnamefont {Van~Assche}},\
  }\href@noop {} {\bibfield  {journal} {\bibinfo  {journal} {Phys. Rev. A}\
  }\textbf {\bibinfo {volume} {63}},\ \bibinfo {pages} {052311} (\bibinfo
  {year} {2001})}\BibitemShut {NoStop}%
\bibitem [{\citenamefont {Grosshans}\ and\ \citenamefont
  {Grangier}(2002)}]{Grosshans2002}%
  \BibitemOpen
  \bibfield  {author} {\bibinfo {author} {\bibfnamefont {F.}~\bibnamefont
  {Grosshans}}\ and\ \bibinfo {author} {\bibfnamefont {P.}~\bibnamefont
  {Grangier}},\ }\href@noop {} {\bibfield  {journal} {\bibinfo  {journal}
  {Phys. Rev. Lett.}\ }\textbf {\bibinfo {volume} {88}},\ \bibinfo {pages}
  {057902} (\bibinfo {year} {2002})}\BibitemShut {NoStop}%
\bibitem [{\citenamefont {Weedbrook}\ \emph {et~al.}(2004)\citenamefont
  {Weedbrook}, \citenamefont {Lance}, \citenamefont {Bowen}, \citenamefont
  {Symul}, \citenamefont {Ralph},\ and\ \citenamefont {Lam}}]{Weedbrook04}%
  \BibitemOpen
  \bibfield  {author} {\bibinfo {author} {\bibfnamefont {C.}~\bibnamefont
  {Weedbrook}}, \bibinfo {author} {\bibfnamefont {A.~M.}\ \bibnamefont
  {Lance}}, \bibinfo {author} {\bibfnamefont {W.~P.}\ \bibnamefont {Bowen}},
  \bibinfo {author} {\bibfnamefont {T.}~\bibnamefont {Symul}}, \bibinfo
  {author} {\bibfnamefont {T.~C.}\ \bibnamefont {Ralph}}, \ and\ \bibinfo
  {author} {\bibfnamefont {P.~K.}\ \bibnamefont {Lam}},\ }\href@noop {}
  {\bibfield  {journal} {\bibinfo  {journal} {Phys. Rev. Lett.}\ }\textbf
  {\bibinfo {volume} {93}},\ \bibinfo {pages} {170504} (\bibinfo {year}
  {2004})}\BibitemShut {NoStop}%
\bibitem [{\citenamefont {Bennett}\ \emph {et~al.}(1992)\citenamefont
  {Bennett}, \citenamefont {Bessette}, \citenamefont {Salvail}, \citenamefont
  {Brassard},\ and\ \citenamefont {Smolin}}]{Bennett92}%
  \BibitemOpen
  \bibfield  {author} {\bibinfo {author} {\bibfnamefont {C.~H.}\ \bibnamefont
  {Bennett}}, \bibinfo {author} {\bibfnamefont {F.}~\bibnamefont {Bessette}},
  \bibinfo {author} {\bibfnamefont {L.}~\bibnamefont {Salvail}}, \bibinfo
  {author} {\bibfnamefont {G.}~\bibnamefont {Brassard}}, \ and\ \bibinfo
  {author} {\bibfnamefont {J.}~\bibnamefont {Smolin}},\ }\href@noop {}
  {\bibfield  {journal} {\bibinfo  {journal} {J. Cryptology}\ }\textbf
  {\bibinfo {volume} {5}},\ \bibinfo {pages} {3} (\bibinfo {year}
  {1992})}\BibitemShut {NoStop}%
\bibitem [{\citenamefont {Muller}\ \emph {et~al.}(1995)\citenamefont {Muller},
  \citenamefont {Zbinden},\ and\ \citenamefont {Gisin}}]{Muller95}%
  \BibitemOpen
  \bibfield  {author} {\bibinfo {author} {\bibfnamefont {A.}~\bibnamefont
  {Muller}}, \bibinfo {author} {\bibfnamefont {H.}~\bibnamefont {Zbinden}}, \
  and\ \bibinfo {author} {\bibfnamefont {N.}~\bibnamefont {Gisin}},\
  }\href@noop {} {\bibfield  {journal} {\bibinfo  {journal} {Nature (London)}\
  }\textbf {\bibinfo {volume} {378}},\ \bibinfo {pages} {449} (\bibinfo {year}
  {1995})}\BibitemShut {NoStop}%
\bibitem [{\citenamefont {Jennewein}\ \emph {et~al.}(2000)\citenamefont
  {Jennewein}, \citenamefont {Simon}, \citenamefont {Weihs}, \citenamefont
  {Weinfurter},\ and\ \citenamefont {Zeilinger}}]{Jennewein00}%
  \BibitemOpen
  \bibfield  {author} {\bibinfo {author} {\bibfnamefont {T.}~\bibnamefont
  {Jennewein}}, \bibinfo {author} {\bibfnamefont {C.}~\bibnamefont {Simon}},
  \bibinfo {author} {\bibfnamefont {G.}~\bibnamefont {Weihs}}, \bibinfo
  {author} {\bibfnamefont {H.}~\bibnamefont {Weinfurter}}, \ and\ \bibinfo
  {author} {\bibfnamefont {A.}~\bibnamefont {Zeilinger}},\ }\href@noop {}
  {\bibfield  {journal} {\bibinfo  {journal} {Phys. Rev. Lett.}\ }\textbf
  {\bibinfo {volume} {84}},\ \bibinfo {pages} {4729} (\bibinfo {year}
  {2000})}\BibitemShut {NoStop}%
\bibitem [{\citenamefont {Naik}\ \emph {et~al.}(2000)\citenamefont {Naik},
  \citenamefont {Peterson}, \citenamefont {White}, \citenamefont {Berglund},\
  and\ \citenamefont {Kwiat}}]{Naik00}%
  \BibitemOpen
  \bibfield  {author} {\bibinfo {author} {\bibfnamefont {D.~S.}\ \bibnamefont
  {Naik}}, \bibinfo {author} {\bibfnamefont {C.~G.}\ \bibnamefont {Peterson}},
  \bibinfo {author} {\bibfnamefont {A.~G.}\ \bibnamefont {White}}, \bibinfo
  {author} {\bibfnamefont {A.~J.}\ \bibnamefont {Berglund}}, \ and\ \bibinfo
  {author} {\bibfnamefont {P.~G.}\ \bibnamefont {Kwiat}},\ }\href@noop {}
  {\bibfield  {journal} {\bibinfo  {journal} {Phys. Rev. Lett.}\ }\textbf
  {\bibinfo {volume} {84}},\ \bibinfo {pages} {4733} (\bibinfo {year}
  {2000})}\BibitemShut {NoStop}%
\bibitem [{\citenamefont {Tittel}\ \emph {et~al.}(2000)\citenamefont {Tittel},
  \citenamefont {Brendel}, \citenamefont {Zbinden},\ and\ \citenamefont
  {Gisin}}]{Tittel00}%
  \BibitemOpen
  \bibfield  {author} {\bibinfo {author} {\bibfnamefont {W.}~\bibnamefont
  {Tittel}}, \bibinfo {author} {\bibfnamefont {J.}~\bibnamefont {Brendel}},
  \bibinfo {author} {\bibfnamefont {H.}~\bibnamefont {Zbinden}}, \ and\
  \bibinfo {author} {\bibfnamefont {N.}~\bibnamefont {Gisin}},\ }\href@noop {}
  {\bibfield  {journal} {\bibinfo  {journal} {Phys. Rev. Lett.}\ }\textbf
  {\bibinfo {volume} {84}},\ \bibinfo {pages} {4737} (\bibinfo {year}
  {2000})}\BibitemShut {NoStop}%
\bibitem [{\citenamefont {Grosshans}\ \emph
  {et~al.}(2003{\natexlab{a}})\citenamefont {Grosshans}, \citenamefont
  {Van~Assche}, \citenamefont {Wenger}, \citenamefont {Brouri}, \citenamefont
  {Cerf},\ and\ \citenamefont {Grangier}}]{Grosshans2003}%
  \BibitemOpen
  \bibfield  {author} {\bibinfo {author} {\bibfnamefont {F.}~\bibnamefont
  {Grosshans}}, \bibinfo {author} {\bibfnamefont {G.}~\bibnamefont
  {Van~Assche}}, \bibinfo {author} {\bibfnamefont {J.}~\bibnamefont {Wenger}},
  \bibinfo {author} {\bibfnamefont {R.}~\bibnamefont {Brouri}}, \bibinfo
  {author} {\bibfnamefont {N.~J.}\ \bibnamefont {Cerf}}, \ and\ \bibinfo
  {author} {\bibfnamefont {P.}~\bibnamefont {Grangier}},\ }\href@noop {}
  {\bibfield  {journal} {\bibinfo  {journal} {Nature (London)}\ }\textbf
  {\bibinfo {volume} {421}},\ \bibinfo {pages} {238} (\bibinfo {year}
  {2003}{\natexlab{a}})}\BibitemShut {NoStop}%
\bibitem [{\citenamefont {Lodewyck}\ \emph {et~al.}(2007)\citenamefont
  {Lodewyck}, \citenamefont {Bloch}, \citenamefont {Garc{\'\i}a-Patr{\'o}n},
  \citenamefont {Fossier}, \citenamefont {Karpov}, \citenamefont {Diamanti},
  \citenamefont {Debuisschert}, \citenamefont {Cerf}, \citenamefont
  {Tualle-Brouri}, \citenamefont {McLaughlin} \emph {et~al.}}]{Lodewyck2007}%
  \BibitemOpen
  \bibfield  {author} {\bibinfo {author} {\bibfnamefont {J.}~\bibnamefont
  {Lodewyck}}, \bibinfo {author} {\bibfnamefont {M.}~\bibnamefont {Bloch}},
  \bibinfo {author} {\bibfnamefont {R.}~\bibnamefont {Garc{\'\i}a-Patr{\'o}n}},
  \bibinfo {author} {\bibfnamefont {S.}~\bibnamefont {Fossier}}, \bibinfo
  {author} {\bibfnamefont {E.}~\bibnamefont {Karpov}}, \bibinfo {author}
  {\bibfnamefont {E.}~\bibnamefont {Diamanti}}, \bibinfo {author}
  {\bibfnamefont {T.}~\bibnamefont {Debuisschert}}, \bibinfo {author}
  {\bibfnamefont {N.~J.}\ \bibnamefont {Cerf}}, \bibinfo {author}
  {\bibfnamefont {R.}~\bibnamefont {Tualle-Brouri}}, \bibinfo {author}
  {\bibfnamefont {S.~W.}\ \bibnamefont {McLaughlin}},  \emph {et~al.},\
  }\href@noop {} {\bibfield  {journal} {\bibinfo  {journal} {Phys. Rev. A}\
  }\textbf {\bibinfo {volume} {76}},\ \bibinfo {pages} {042305} (\bibinfo
  {year} {2007})}\BibitemShut {NoStop}%
\bibitem [{\citenamefont {Huang}\ \emph {et~al.}(2016)\citenamefont {Huang},
  \citenamefont {Huang}, \citenamefont {Lin},\ and\ \citenamefont
  {Zeng}}]{Huang2016}%
  \BibitemOpen
  \bibfield  {author} {\bibinfo {author} {\bibfnamefont {D.}~\bibnamefont
  {Huang}}, \bibinfo {author} {\bibfnamefont {P.}~\bibnamefont {Huang}},
  \bibinfo {author} {\bibfnamefont {D.}~\bibnamefont {Lin}}, \ and\ \bibinfo
  {author} {\bibfnamefont {G.}~\bibnamefont {Zeng}},\ }\href@noop {} {\bibfield
   {journal} {\bibinfo  {journal} {Sci. Rep.}\ }\textbf {\bibinfo {volume}
  {6}},\ \bibinfo {pages} {19201} (\bibinfo {year} {2016})}\BibitemShut
  {NoStop}%
\bibitem [{\citenamefont {Madsen}\ \emph {et~al.}(2012)\citenamefont {Madsen},
  \citenamefont {Usenko}, \citenamefont {Lassen}, \citenamefont {Filip},\ and\
  \citenamefont {Andersen}}]{Madsen2012}%
  \BibitemOpen
  \bibfield  {author} {\bibinfo {author} {\bibfnamefont {L.~S.}\ \bibnamefont
  {Madsen}}, \bibinfo {author} {\bibfnamefont {V.~C.}\ \bibnamefont {Usenko}},
  \bibinfo {author} {\bibfnamefont {M.}~\bibnamefont {Lassen}}, \bibinfo
  {author} {\bibfnamefont {R.}~\bibnamefont {Filip}}, \ and\ \bibinfo {author}
  {\bibfnamefont {U.~L.}\ \bibnamefont {Andersen}},\ }\href@noop {} {\bibfield
  {journal} {\bibinfo  {journal} {Nat. Commun.}\ }\textbf {\bibinfo {volume}
  {3}},\ \bibinfo {pages} {1083} (\bibinfo {year} {2012})}\BibitemShut
  {NoStop}%
\bibitem [{\citenamefont {Jouguet}\ \emph {et~al.}(2013)\citenamefont
  {Jouguet}, \citenamefont {Kunz-Jacques}, \citenamefont {Leverrier},
  \citenamefont {Grangier},\ and\ \citenamefont {Diamanti}}]{Jouguet2013}%
  \BibitemOpen
  \bibfield  {author} {\bibinfo {author} {\bibfnamefont {P.}~\bibnamefont
  {Jouguet}}, \bibinfo {author} {\bibfnamefont {S.}~\bibnamefont
  {Kunz-Jacques}}, \bibinfo {author} {\bibfnamefont {A.}~\bibnamefont
  {Leverrier}}, \bibinfo {author} {\bibfnamefont {P.}~\bibnamefont {Grangier}},
  \ and\ \bibinfo {author} {\bibfnamefont {E.}~\bibnamefont {Diamanti}},\
  }\href@noop {} {\bibfield  {journal} {\bibinfo  {journal} {Nat. Photon.}\
  }\textbf {\bibinfo {volume} {7}},\ \bibinfo {pages} {378} (\bibinfo {year}
  {2013})}\BibitemShut {NoStop}%
\bibitem [{\citenamefont {L\"{u}tkenhaus}(1996)}]{Lutkenhaus96}%
  \BibitemOpen
  \bibfield  {author} {\bibinfo {author} {\bibfnamefont {N.}~\bibnamefont
  {L\"{u}tkenhaus}},\ }\href@noop {} {\bibfield  {journal} {\bibinfo  {journal}
  {Phys. Rev. A}\ }\textbf {\bibinfo {volume} {54}},\ \bibinfo {pages} {97}
  (\bibinfo {year} {1996})}\BibitemShut {NoStop}%
\bibitem [{\citenamefont {Slutsky}\ \emph {et~al.}(1998)\citenamefont
  {Slutsky}, \citenamefont {Rao}, \citenamefont {Sun},\ and\ \citenamefont
  {Fainman}}]{Slutsky98}%
  \BibitemOpen
  \bibfield  {author} {\bibinfo {author} {\bibfnamefont {B.~A.}\ \bibnamefont
  {Slutsky}}, \bibinfo {author} {\bibfnamefont {R.}~\bibnamefont {Rao}},
  \bibinfo {author} {\bibfnamefont {P.-C.}\ \bibnamefont {Sun}}, \ and\
  \bibinfo {author} {\bibfnamefont {Y.}~\bibnamefont {Fainman}},\ }\href@noop
  {} {\bibfield  {journal} {\bibinfo  {journal} {Phys. Rev. A}\ }\textbf
  {\bibinfo {volume} {57}},\ \bibinfo {pages} {2383} (\bibinfo {year}
  {1998})}\BibitemShut {NoStop}%
\bibitem [{\citenamefont {Bechmann-Pasquinucci}(2006)}]{Bechmann06}%
  \BibitemOpen
  \bibfield  {author} {\bibinfo {author} {\bibfnamefont {H.}~\bibnamefont
  {Bechmann-Pasquinucci}},\ }\href@noop {} {\bibfield  {journal} {\bibinfo
  {journal} {Phys. Rev. A}\ }\textbf {\bibinfo {volume} {73}},\ \bibinfo
  {pages} {044305} (\bibinfo {year} {2006})}\BibitemShut {NoStop}%
\bibitem [{\citenamefont {Grosshans}\ \emph
  {et~al.}(2003{\natexlab{b}})\citenamefont {Grosshans}, \citenamefont {Cerf},
  \citenamefont {Wenger}, \citenamefont {Tualle-Brouri},\ and\ \citenamefont
  {Grangier}}]{Grosshans2003a}%
  \BibitemOpen
  \bibfield  {author} {\bibinfo {author} {\bibfnamefont {F.}~\bibnamefont
  {Grosshans}}, \bibinfo {author} {\bibfnamefont {N.~J.}\ \bibnamefont {Cerf}},
  \bibinfo {author} {\bibfnamefont {J.}~\bibnamefont {Wenger}}, \bibinfo
  {author} {\bibfnamefont {R.}~\bibnamefont {Tualle-Brouri}}, \ and\ \bibinfo
  {author} {\bibfnamefont {P.}~\bibnamefont {Grangier}},\ }\href@noop {}
  {\bibfield  {journal} {\bibinfo  {journal} {Quant. Inf. Comput.}\ }\textbf
  {\bibinfo {volume} {3}},\ \bibinfo {pages} {535} (\bibinfo {year}
  {2003}{\natexlab{b}})}\BibitemShut {NoStop}%
\bibitem [{\citenamefont {Biham}\ and\ \citenamefont {Mor}(1997)}]{Biham97}%
  \BibitemOpen
  \bibfield  {author} {\bibinfo {author} {\bibfnamefont {E.}~\bibnamefont
  {Biham}}\ and\ \bibinfo {author} {\bibfnamefont {T.}~\bibnamefont {Mor}},\
  }\href@noop {} {\bibfield  {journal} {\bibinfo  {journal} {Phys. Rev. Lett.}\
  }\textbf {\bibinfo {volume} {78}},\ \bibinfo {pages} {2256} (\bibinfo {year}
  {1997})}\BibitemShut {NoStop}%
\bibitem [{\citenamefont {Biham}\ \emph {et~al.}(2002)\citenamefont {Biham},
  \citenamefont {Boyer}, \citenamefont {Brassard}, \citenamefont {{van de
  Graaf}},\ and\ \citenamefont {Mor}}]{Biham02}%
  \BibitemOpen
  \bibfield  {author} {\bibinfo {author} {\bibfnamefont {E.}~\bibnamefont
  {Biham}}, \bibinfo {author} {\bibfnamefont {M.}~\bibnamefont {Boyer}},
  \bibinfo {author} {\bibfnamefont {G.}~\bibnamefont {Brassard}}, \bibinfo
  {author} {\bibfnamefont {J.}~\bibnamefont {{van de Graaf}}}, \ and\ \bibinfo
  {author} {\bibfnamefont {T.}~\bibnamefont {Mor}},\ }\href@noop {} {\bibfield
  {journal} {\bibinfo  {journal} {Algorithmica}\ }\textbf {\bibinfo {volume}
  {34}},\ \bibinfo {pages} {372} (\bibinfo {year} {2002})}\BibitemShut
  {NoStop}%
\bibitem [{\citenamefont {Navascu{\'e}s}\ \emph {et~al.}(2006)\citenamefont
  {Navascu{\'e}s}, \citenamefont {Grosshans},\ and\ \citenamefont
  {Acin}}]{Navascues2006}%
  \BibitemOpen
  \bibfield  {author} {\bibinfo {author} {\bibfnamefont {M.}~\bibnamefont
  {Navascu{\'e}s}}, \bibinfo {author} {\bibfnamefont {F.}~\bibnamefont
  {Grosshans}}, \ and\ \bibinfo {author} {\bibfnamefont {A.}~\bibnamefont
  {Acin}},\ }\href@noop {} {\bibfield  {journal} {\bibinfo  {journal} {Phys.
  Rev. Lett.}\ }\textbf {\bibinfo {volume} {97}},\ \bibinfo {pages} {190502}
  (\bibinfo {year} {2006})}\BibitemShut {NoStop}%
\bibitem [{\citenamefont {Garc\'{i}a-Patr\'{o}n}\ and\ \citenamefont
  {Cerf}(2006)}]{Garcia2006}%
  \BibitemOpen
  \bibfield  {author} {\bibinfo {author} {\bibfnamefont {R.}~\bibnamefont
  {Garc\'{i}a-Patr\'{o}n}}\ and\ \bibinfo {author} {\bibfnamefont {N.~J.}\
  \bibnamefont {Cerf}},\ }\href@noop {} {\bibfield  {journal} {\bibinfo
  {journal} {Phys. Rev. Lett.}\ }\textbf {\bibinfo {volume} {97}},\ \bibinfo
  {pages} {190503} (\bibinfo {year} {2006})}\BibitemShut {NoStop}%
\bibitem [{\citenamefont {Kraus}\ \emph {et~al.}(2005)\citenamefont {Kraus},
  \citenamefont {Gisin},\ and\ \citenamefont {Renner}}]{Kraus05}%
  \BibitemOpen
  \bibfield  {author} {\bibinfo {author} {\bibfnamefont {B.}~\bibnamefont
  {Kraus}}, \bibinfo {author} {\bibfnamefont {N.}~\bibnamefont {Gisin}}, \ and\
  \bibinfo {author} {\bibfnamefont {R.}~\bibnamefont {Renner}},\ }\href@noop {}
  {\bibfield  {journal} {\bibinfo  {journal} {Phys. Rev. Lett.}\ }\textbf
  {\bibinfo {volume} {95}},\ \bibinfo {pages} {080501} (\bibinfo {year}
  {2005})}\BibitemShut {NoStop}%
\bibitem [{\citenamefont {Renner}\ \emph {et~al.}(2005)\citenamefont {Renner},
  \citenamefont {Gisin},\ and\ \citenamefont {Kraus}}]{Renner05}%
  \BibitemOpen
  \bibfield  {author} {\bibinfo {author} {\bibfnamefont {R.}~\bibnamefont
  {Renner}}, \bibinfo {author} {\bibfnamefont {N.}~\bibnamefont {Gisin}}, \
  and\ \bibinfo {author} {\bibfnamefont {B.}~\bibnamefont {Kraus}},\
  }\href@noop {} {\bibfield  {journal} {\bibinfo  {journal} {Phys. Rev. A}\
  }\textbf {\bibinfo {volume} {72}},\ \bibinfo {pages} {012332} (\bibinfo
  {year} {2005})}\BibitemShut {NoStop}%
\bibitem [{\citenamefont {Leverrier}\ \emph {et~al.}(2013)\citenamefont
  {Leverrier}, \citenamefont {Garc{\'\i}a-Patr{\'o}n}, \citenamefont {Renner},\
  and\ \citenamefont {Cerf}}]{Leverrier2013}%
  \BibitemOpen
  \bibfield  {author} {\bibinfo {author} {\bibfnamefont {A.}~\bibnamefont
  {Leverrier}}, \bibinfo {author} {\bibfnamefont {R.}~\bibnamefont
  {Garc{\'\i}a-Patr{\'o}n}}, \bibinfo {author} {\bibfnamefont {R.}~\bibnamefont
  {Renner}}, \ and\ \bibinfo {author} {\bibfnamefont {N.~J.}\ \bibnamefont
  {Cerf}},\ }\href@noop {} {\bibfield  {journal} {\bibinfo  {journal} {Phys.
  Rev. Lett.}\ }\textbf {\bibinfo {volume} {110}},\ \bibinfo {pages} {030502}
  (\bibinfo {year} {2013})}\BibitemShut {NoStop}%
\bibitem [{\citenamefont {Hasegawa}\ \emph {et~al.}(2007)\citenamefont
  {Hasegawa}, \citenamefont {Hayashi}, \citenamefont {Hiroshima},\ and\
  \citenamefont {Tomita}}]{Hasegawa07}%
  \BibitemOpen
  \bibfield  {author} {\bibinfo {author} {\bibfnamefont {J.}~\bibnamefont
  {Hasegawa}}, \bibinfo {author} {\bibfnamefont {M.}~\bibnamefont {Hayashi}},
  \bibinfo {author} {\bibfnamefont {T.}~\bibnamefont {Hiroshima}}, \ and\
  \bibinfo {author} {\bibfnamefont {A.}~\bibnamefont {Tomita}},\ }\href@noop {}
  {\bibfield  {journal} {\bibinfo  {journal} {arXiv:0707.3541}\ } (\bibinfo
  {year} {2007})}\BibitemShut {NoStop}%
\bibitem [{\citenamefont {Hayashi}(2007)}]{Hayashi07}%
  \BibitemOpen
  \bibfield  {author} {\bibinfo {author} {\bibfnamefont {M.}~\bibnamefont
  {Hayashi}},\ }\href@noop {} {\bibfield  {journal} {\bibinfo  {journal} {Phys.
  Rev. A}\ }\textbf {\bibinfo {volume} {76}},\ \bibinfo {pages} {012329}
  (\bibinfo {year} {2007})}\BibitemShut {NoStop}%
\bibitem [{\citenamefont {Scarani}\ and\ \citenamefont
  {Renner}(2008)}]{Scarani08}%
  \BibitemOpen
  \bibfield  {author} {\bibinfo {author} {\bibfnamefont {V.}~\bibnamefont
  {Scarani}}\ and\ \bibinfo {author} {\bibfnamefont {R.}~\bibnamefont
  {Renner}},\ }\href@noop {} {\bibfield  {journal} {\bibinfo  {journal} {Phys.
  Rev. Lett.}\ }\textbf {\bibinfo {volume} {100}},\ \bibinfo {pages} {200501}
  (\bibinfo {year} {2008})}\BibitemShut {NoStop}%
\bibitem [{\citenamefont {Leverrier}\ \emph {et~al.}(2010)\citenamefont
  {Leverrier}, \citenamefont {Grosshans},\ and\ \citenamefont
  {Grangier}}]{Leverrier2010}%
  \BibitemOpen
  \bibfield  {author} {\bibinfo {author} {\bibfnamefont {A.}~\bibnamefont
  {Leverrier}}, \bibinfo {author} {\bibfnamefont {F.}~\bibnamefont
  {Grosshans}}, \ and\ \bibinfo {author} {\bibfnamefont {P.}~\bibnamefont
  {Grangier}},\ }\href@noop {} {\bibfield  {journal} {\bibinfo  {journal}
  {Phys. Rev. A}\ }\textbf {\bibinfo {volume} {81}},\ \bibinfo {pages} {062343}
  (\bibinfo {year} {2010})}\BibitemShut {NoStop}%
\bibitem [{\citenamefont {Ruppert}\ \emph {et~al.}(2014)\citenamefont
  {Ruppert}, \citenamefont {Usenko},\ and\ \citenamefont
  {Filip}}]{Ruppert2014}%
  \BibitemOpen
  \bibfield  {author} {\bibinfo {author} {\bibfnamefont {L.}~\bibnamefont
  {Ruppert}}, \bibinfo {author} {\bibfnamefont {V.~C.}\ \bibnamefont {Usenko}},
  \ and\ \bibinfo {author} {\bibfnamefont {R.}~\bibnamefont {Filip}},\
  }\href@noop {} {\bibfield  {journal} {\bibinfo  {journal} {Phys. Rev. A}\
  }\textbf {\bibinfo {volume} {90}},\ \bibinfo {pages} {062310} (\bibinfo
  {year} {2014})}\BibitemShut {NoStop}%
\bibitem [{\citenamefont {L\"{u}tkenhaus}(1999)}]{Lutkenhaus99}%
  \BibitemOpen
  \bibfield  {author} {\bibinfo {author} {\bibfnamefont {N.}~\bibnamefont
  {L\"{u}tkenhaus}},\ }\href@noop {} {\bibfield  {journal} {\bibinfo  {journal}
  {Phys. Rev. A}\ }\textbf {\bibinfo {volume} {59}},\ \bibinfo {pages} {3301}
  (\bibinfo {year} {1999})}\BibitemShut {NoStop}%
\bibitem [{\citenamefont {Brassard}\ \emph {et~al.}(2000)\citenamefont
  {Brassard}, \citenamefont {L\"{u}tkenhaus}, \citenamefont {Mor},\ and\
  \citenamefont {Sanders}}]{Brassard00}%
  \BibitemOpen
  \bibfield  {author} {\bibinfo {author} {\bibfnamefont {G.}~\bibnamefont
  {Brassard}}, \bibinfo {author} {\bibfnamefont {N.}~\bibnamefont
  {L\"{u}tkenhaus}}, \bibinfo {author} {\bibfnamefont {T.}~\bibnamefont {Mor}},
  \ and\ \bibinfo {author} {\bibfnamefont {B.~C.}\ \bibnamefont {Sanders}},\
  }\href@noop {} {\bibfield  {journal} {\bibinfo  {journal} {Phys. Rev. Lett.}\
  }\textbf {\bibinfo {volume} {85}},\ \bibinfo {pages} {1330} (\bibinfo {year}
  {2000})}\BibitemShut {NoStop}%
\bibitem [{\citenamefont {Gottesman}\ \emph {et~al.}(2004)\citenamefont
  {Gottesman}, \citenamefont {Lo}, \citenamefont {L\"{u}tkenhaus},\ and\
  \citenamefont {Preskill}}]{Gottesman04}%
  \BibitemOpen
  \bibfield  {author} {\bibinfo {author} {\bibfnamefont {D.}~\bibnamefont
  {Gottesman}}, \bibinfo {author} {\bibfnamefont {H.-K.}\ \bibnamefont {Lo}},
  \bibinfo {author} {\bibfnamefont {N.}~\bibnamefont {L\"{u}tkenhaus}}, \ and\
  \bibinfo {author} {\bibfnamefont {J.}~\bibnamefont {Preskill}},\ }\href@noop
  {} {\bibfield  {journal} {\bibinfo  {journal} {Quant. Inf. Comput.}\ }\textbf
  {\bibinfo {volume} {5}},\ \bibinfo {pages} {325} (\bibinfo {year}
  {2004})}\BibitemShut {NoStop}%
\bibitem [{\citenamefont {Filip}(2008)}]{Filip2008}%
  \BibitemOpen
  \bibfield  {author} {\bibinfo {author} {\bibfnamefont {R.}~\bibnamefont
  {Filip}},\ }\href@noop {} {\bibfield  {journal} {\bibinfo  {journal} {Phys.
  Rev. A}\ }\textbf {\bibinfo {volume} {77}},\ \bibinfo {pages} {022310}
  (\bibinfo {year} {2008})}\BibitemShut {NoStop}%
\bibitem [{\citenamefont {Usenko}\ and\ \citenamefont
  {Filip}(2010)}]{Usenko2010a}%
  \BibitemOpen
  \bibfield  {author} {\bibinfo {author} {\bibfnamefont {V.~C.}\ \bibnamefont
  {Usenko}}\ and\ \bibinfo {author} {\bibfnamefont {R.}~\bibnamefont {Filip}},\
  }\href@noop {} {\bibfield  {journal} {\bibinfo  {journal} {Phys. Rev. A}\
  }\textbf {\bibinfo {volume} {81}},\ \bibinfo {pages} {022318} (\bibinfo
  {year} {2010})}\BibitemShut {NoStop}%
\bibitem [{\citenamefont {Jouguet}\ \emph {et~al.}(2012)\citenamefont
  {Jouguet}, \citenamefont {Kunz-Jacques}, \citenamefont {Diamanti},\ and\
  \citenamefont {Leverrier}}]{Jouguet2012}%
  \BibitemOpen
  \bibfield  {author} {\bibinfo {author} {\bibfnamefont {P.}~\bibnamefont
  {Jouguet}}, \bibinfo {author} {\bibfnamefont {S.}~\bibnamefont
  {Kunz-Jacques}}, \bibinfo {author} {\bibfnamefont {E.}~\bibnamefont
  {Diamanti}}, \ and\ \bibinfo {author} {\bibfnamefont {A.}~\bibnamefont
  {Leverrier}},\ }\href@noop {} {\bibfield  {journal} {\bibinfo  {journal}
  {Phys. Rev. A}\ }\textbf {\bibinfo {volume} {86}},\ \bibinfo {pages} {032309}
  (\bibinfo {year} {2012})}\BibitemShut {NoStop}%
\bibitem [{\citenamefont {Usenko}\ and\ \citenamefont
  {Filip}(2016)}]{Usenko2016}%
  \BibitemOpen
  \bibfield  {author} {\bibinfo {author} {\bibfnamefont {V.~C.}\ \bibnamefont
  {Usenko}}\ and\ \bibinfo {author} {\bibfnamefont {R.}~\bibnamefont {Filip}},\
  }\href@noop {} {\bibfield  {journal} {\bibinfo  {journal} {Entropy}\ }\textbf
  {\bibinfo {volume} {18}},\ \bibinfo {pages} {20} (\bibinfo {year}
  {2016})}\BibitemShut {NoStop}%
\bibitem [{\citenamefont {Garc{\'\i}a-Patr{\'o}n}\ and\ \citenamefont
  {Cerf}(2009)}]{Garcia2009}%
  \BibitemOpen
  \bibfield  {author} {\bibinfo {author} {\bibfnamefont {R.}~\bibnamefont
  {Garc{\'\i}a-Patr{\'o}n}}\ and\ \bibinfo {author} {\bibfnamefont {N.~J.}\
  \bibnamefont {Cerf}},\ }\href@noop {} {\bibfield  {journal} {\bibinfo
  {journal} {Phys. Rev. Lett.}\ }\textbf {\bibinfo {volume} {102}},\ \bibinfo
  {pages} {130501} (\bibinfo {year} {2009})}\BibitemShut {NoStop}%
\bibitem [{\citenamefont {Xu}\ \emph {et~al.}(2015)\citenamefont {Xu},
  \citenamefont {Curty}, \citenamefont {Qi}, \citenamefont {Qian},\ and\
  \citenamefont {Lo}}]{Xu15}%
  \BibitemOpen
  \bibfield  {author} {\bibinfo {author} {\bibfnamefont {F.}~\bibnamefont
  {Xu}}, \bibinfo {author} {\bibfnamefont {M.}~\bibnamefont {Curty}}, \bibinfo
  {author} {\bibfnamefont {B.}~\bibnamefont {Qi}}, \bibinfo {author}
  {\bibfnamefont {L.}~\bibnamefont {Qian}}, \ and\ \bibinfo {author}
  {\bibfnamefont {H.-K.}\ \bibnamefont {Lo}},\ }\href@noop {} {\bibfield
  {journal} {\bibinfo  {journal} {Nat. Photon.}\ }\textbf {\bibinfo {volume}
  {9}},\ \bibinfo {pages} {772} (\bibinfo {year} {2015})}\BibitemShut {NoStop}%
\bibitem [{\citenamefont {Pirandola}\ \emph {et~al.}(2015)\citenamefont
  {Pirandola}, \citenamefont {Ottaviani}, \citenamefont {Spedalieri},
  \citenamefont {Weedbrook}, \citenamefont {Braunstein}, \citenamefont {Lloyd},
  \citenamefont {Gehring}, \citenamefont {Jacobsen},\ and\ \citenamefont
  {Andersen}}]{Pirandola15}%
  \BibitemOpen
  \bibfield  {author} {\bibinfo {author} {\bibfnamefont {S.}~\bibnamefont
  {Pirandola}}, \bibinfo {author} {\bibfnamefont {C.}~\bibnamefont
  {Ottaviani}}, \bibinfo {author} {\bibfnamefont {G.}~\bibnamefont
  {Spedalieri}}, \bibinfo {author} {\bibfnamefont {C.}~\bibnamefont
  {Weedbrook}}, \bibinfo {author} {\bibfnamefont {S.~L.}\ \bibnamefont
  {Braunstein}}, \bibinfo {author} {\bibfnamefont {S.}~\bibnamefont {Lloyd}},
  \bibinfo {author} {\bibfnamefont {T.}~\bibnamefont {Gehring}}, \bibinfo
  {author} {\bibfnamefont {C.~S.}\ \bibnamefont {Jacobsen}}, \ and\ \bibinfo
  {author} {\bibfnamefont {U.~L.}\ \bibnamefont {Andersen}},\ }\href@noop {}
  {\bibfield  {journal} {\bibinfo  {journal} {Nat. Photon.}\ }\textbf {\bibinfo
  {volume} {9}},\ \bibinfo {pages} {773} (\bibinfo {year} {2015})}\BibitemShut
  {NoStop}%
\bibitem [{\citenamefont {Castelletto}\ \emph {et~al.}(2003)\citenamefont
  {Castelletto}, \citenamefont {Degiovanni},\ and\ \citenamefont
  {Rastello}}]{Castelletto03}%
  \BibitemOpen
  \bibfield  {author} {\bibinfo {author} {\bibfnamefont {S.}~\bibnamefont
  {Castelletto}}, \bibinfo {author} {\bibfnamefont {I.~P.}\ \bibnamefont
  {Degiovanni}}, \ and\ \bibinfo {author} {\bibfnamefont {M.~L.}\ \bibnamefont
  {Rastello}},\ }\href@noop {} {\bibfield  {journal} {\bibinfo  {journal}
  {Phys. Rev. A}\ }\textbf {\bibinfo {volume} {67}},\ \bibinfo {pages} {022305}
  (\bibinfo {year} {2003})}\BibitemShut {NoStop}%
\bibitem [{\citenamefont {Dong}\ \emph {et~al.}(2011)\citenamefont {Dong},
  \citenamefont {Xiu}, \citenamefont {Gao},\ and\ \citenamefont {Yi}}]{Dong11}%
  \BibitemOpen
  \bibfield  {author} {\bibinfo {author} {\bibfnamefont {L.}~\bibnamefont
  {Dong}}, \bibinfo {author} {\bibfnamefont {X.-M.}\ \bibnamefont {Xiu}},
  \bibinfo {author} {\bibfnamefont {Y.-J.}\ \bibnamefont {Gao}}, \ and\
  \bibinfo {author} {\bibfnamefont {X.~X.}\ \bibnamefont {Yi}},\ }\href@noop {}
  {\bibfield  {journal} {\bibinfo  {journal} {J. Exp. Theor. Phys.}\ }\textbf
  {\bibinfo {volume} {113}},\ \bibinfo {pages} {583} (\bibinfo {year}
  {2011})}\BibitemShut {NoStop}%
\bibitem [{\citenamefont {Miao}\ \emph {et~al.}(2005)\citenamefont {Miao},
  \citenamefont {Han}, \citenamefont {Gong}, \citenamefont {Zhang},
  \citenamefont {Diao},\ and\ \citenamefont {Guo}}]{Miao05}%
  \BibitemOpen
  \bibfield  {author} {\bibinfo {author} {\bibfnamefont {E.-L.}\ \bibnamefont
  {Miao}}, \bibinfo {author} {\bibfnamefont {Z.-F.}\ \bibnamefont {Han}},
  \bibinfo {author} {\bibfnamefont {S.-S.}\ \bibnamefont {Gong}}, \bibinfo
  {author} {\bibfnamefont {T.}~\bibnamefont {Zhang}}, \bibinfo {author}
  {\bibfnamefont {D.-S.}\ \bibnamefont {Diao}}, \ and\ \bibinfo {author}
  {\bibfnamefont {G.-C.}\ \bibnamefont {Guo}},\ }\href@noop {} {\bibfield
  {journal} {\bibinfo  {journal} {New J. Phys.}\ }\textbf {\bibinfo {volume}
  {7}},\ \bibinfo {pages} {215} (\bibinfo {year} {2005})}\BibitemShut {NoStop}%
\bibitem [{\citenamefont {Bonato}\ \emph {et~al.}(2009)\citenamefont {Bonato},
  \citenamefont {Tomaello}, \citenamefont {Deppo}, \citenamefont {Naletto},\
  and\ \citenamefont {Villoresi}}]{Bonato09}%
  \BibitemOpen
  \bibfield  {author} {\bibinfo {author} {\bibfnamefont {C.}~\bibnamefont
  {Bonato}}, \bibinfo {author} {\bibfnamefont {A.}~\bibnamefont {Tomaello}},
  \bibinfo {author} {\bibfnamefont {V.~D.}\ \bibnamefont {Deppo}}, \bibinfo
  {author} {\bibfnamefont {G.}~\bibnamefont {Naletto}}, \ and\ \bibinfo
  {author} {\bibfnamefont {P.}~\bibnamefont {Villoresi}},\ }\href@noop {}
  {\bibfield  {journal} {\bibinfo  {journal} {New J. Phys.}\ }\textbf {\bibinfo
  {volume} {11}},\ \bibinfo {pages} {045017} (\bibinfo {year}
  {2009})}\BibitemShut {NoStop}%
\bibitem [{\citenamefont {Bourgoin}\ \emph {et~al.}(2013)\citenamefont
  {Bourgoin}, \citenamefont {Meyer-Scott}, \citenamefont {Higgins},
  \citenamefont {Helou}, \citenamefont {Erven}, \citenamefont {H.H\"{u}bel},
  \citenamefont {Kumar}, \citenamefont {Hudson}, \citenamefont {D'Souza},
  \citenamefont {Girard}, \citenamefont {Laflamme},\ and\ \citenamefont
  {Jennewein}}]{Bourgoin13}%
  \BibitemOpen
  \bibfield  {author} {\bibinfo {author} {\bibfnamefont {J.-P.}\ \bibnamefont
  {Bourgoin}}, \bibinfo {author} {\bibfnamefont {E.}~\bibnamefont
  {Meyer-Scott}}, \bibinfo {author} {\bibfnamefont {B.~L.}\ \bibnamefont
  {Higgins}}, \bibinfo {author} {\bibfnamefont {B.}~\bibnamefont {Helou}},
  \bibinfo {author} {\bibfnamefont {C.}~\bibnamefont {Erven}}, \bibinfo
  {author} {\bibnamefont {H.H\"{u}bel}}, \bibinfo {author} {\bibfnamefont
  {B.}~\bibnamefont {Kumar}}, \bibinfo {author} {\bibfnamefont
  {D.}~\bibnamefont {Hudson}}, \bibinfo {author} {\bibfnamefont
  {I.}~\bibnamefont {D'Souza}}, \bibinfo {author} {\bibfnamefont
  {R.}~\bibnamefont {Girard}}, \bibinfo {author} {\bibfnamefont
  {R.}~\bibnamefont {Laflamme}}, \ and\ \bibinfo {author} {\bibfnamefont
  {T.}~\bibnamefont {Jennewein}},\ }\href@noop {} {\bibfield  {journal}
  {\bibinfo  {journal} {New J. Phys.}\ }\textbf {\bibinfo {volume} {15}},\
  \bibinfo {pages} {023006} (\bibinfo {year} {2013})}\BibitemShut {NoStop}%
\bibitem [{\citenamefont {Muller}\ \emph {et~al.}(1997)\citenamefont {Muller},
  \citenamefont {Herzog}, \citenamefont {Huttner}, \citenamefont {Tittel},
  \citenamefont {Zbinden},\ and\ \citenamefont {Gisin}}]{Muller97}%
  \BibitemOpen
  \bibfield  {author} {\bibinfo {author} {\bibfnamefont {A.}~\bibnamefont
  {Muller}}, \bibinfo {author} {\bibfnamefont {T.}~\bibnamefont {Herzog}},
  \bibinfo {author} {\bibfnamefont {B.}~\bibnamefont {Huttner}}, \bibinfo
  {author} {\bibfnamefont {W.}~\bibnamefont {Tittel}}, \bibinfo {author}
  {\bibfnamefont {H.}~\bibnamefont {Zbinden}}, \ and\ \bibinfo {author}
  {\bibfnamefont {N.}~\bibnamefont {Gisin}},\ }\href@noop {} {\bibfield
  {journal} {\bibinfo  {journal} {Appl. Phys. Lett.}\ }\textbf {\bibinfo
  {volume} {70}},\ \bibinfo {pages} {793} (\bibinfo {year} {1997})}\BibitemShut
  {NoStop}%
\bibitem [{\citenamefont {Stucki}\ \emph {et~al.}(2002)\citenamefont {Stucki},
  \citenamefont {Gisin}, \citenamefont {Guinnard}, \citenamefont {Ribordy},\
  and\ \citenamefont {Zbinden}}]{Stucki02}%
  \BibitemOpen
  \bibfield  {author} {\bibinfo {author} {\bibfnamefont {D.}~\bibnamefont
  {Stucki}}, \bibinfo {author} {\bibfnamefont {N.}~\bibnamefont {Gisin}},
  \bibinfo {author} {\bibfnamefont {O.}~\bibnamefont {Guinnard}}, \bibinfo
  {author} {\bibfnamefont {G.}~\bibnamefont {Ribordy}}, \ and\ \bibinfo
  {author} {\bibfnamefont {H.}~\bibnamefont {Zbinden}},\ }\href@noop {}
  {\bibfield  {journal} {\bibinfo  {journal} {New J. Phys.}\ }\textbf {\bibinfo
  {volume} {4}},\ \bibinfo {pages} {41} (\bibinfo {year} {2002})}\BibitemShut
  {NoStop}%
\bibitem [{\citenamefont {Zanardi}\ and\ \citenamefont
  {Rasetti}(1997)}]{Zanardi97}%
  \BibitemOpen
  \bibfield  {author} {\bibinfo {author} {\bibfnamefont {P.}~\bibnamefont
  {Zanardi}}\ and\ \bibinfo {author} {\bibfnamefont {M.}~\bibnamefont
  {Rasetti}},\ }\href@noop {} {\bibfield  {journal} {\bibinfo  {journal} {Phys.
  Rev. Lett.}\ }\textbf {\bibinfo {volume} {79}},\ \bibinfo {pages} {3306}
  (\bibinfo {year} {1997})}\BibitemShut {NoStop}%
\bibitem [{\citenamefont {Kempe}\ \emph {et~al.}(2001)\citenamefont {Kempe},
  \citenamefont {Bacon}, \citenamefont {Lidar},\ and\ \citenamefont
  {Whaley}}]{Kempe01}%
  \BibitemOpen
  \bibfield  {author} {\bibinfo {author} {\bibfnamefont {J.}~\bibnamefont
  {Kempe}}, \bibinfo {author} {\bibfnamefont {D.}~\bibnamefont {Bacon}},
  \bibinfo {author} {\bibfnamefont {D.~A.}\ \bibnamefont {Lidar}}, \ and\
  \bibinfo {author} {\bibfnamefont {K.~B.}\ \bibnamefont {Whaley}},\
  }\href@noop {} {\bibfield  {journal} {\bibinfo  {journal} {Phys. Rev. A}\
  }\textbf {\bibinfo {volume} {63}},\ \bibinfo {pages} {042307} (\bibinfo
  {year} {2001})}\BibitemShut {NoStop}%
\bibitem [{\citenamefont {Walton}\ \emph {et~al.}(2003)\citenamefont {Walton},
  \citenamefont {Abouraddy}, \citenamefont {Sergienko}, \citenamefont {Saleh},\
  and\ \citenamefont {Teich}}]{Walton03}%
  \BibitemOpen
  \bibfield  {author} {\bibinfo {author} {\bibfnamefont {Z.~D.}\ \bibnamefont
  {Walton}}, \bibinfo {author} {\bibfnamefont {A.~F.}\ \bibnamefont
  {Abouraddy}}, \bibinfo {author} {\bibfnamefont {A.~V.}\ \bibnamefont
  {Sergienko}}, \bibinfo {author} {\bibfnamefont {B.~E.~A.}\ \bibnamefont
  {Saleh}}, \ and\ \bibinfo {author} {\bibfnamefont {M.~C.}\ \bibnamefont
  {Teich}},\ }\href@noop {} {\bibfield  {journal} {\bibinfo  {journal} {Phys.
  Rev. Lett.}\ }\textbf {\bibinfo {volume} {91}},\ \bibinfo {pages} {087901}
  (\bibinfo {year} {2003})}\BibitemShut {NoStop}%
\bibitem [{\citenamefont {Boileau}\ \emph {et~al.}(2004)\citenamefont
  {Boileau}, \citenamefont {Gottesman}, \citenamefont {Laflamme}, \citenamefont
  {Poulin},\ and\ \citenamefont {Spekkens}}]{Boileau04}%
  \BibitemOpen
  \bibfield  {author} {\bibinfo {author} {\bibfnamefont {J.-C.}\ \bibnamefont
  {Boileau}}, \bibinfo {author} {\bibfnamefont {D.}~\bibnamefont {Gottesman}},
  \bibinfo {author} {\bibfnamefont {R.}~\bibnamefont {Laflamme}}, \bibinfo
  {author} {\bibfnamefont {D.}~\bibnamefont {Poulin}}, \ and\ \bibinfo {author}
  {\bibfnamefont {R.~W.}\ \bibnamefont {Spekkens}},\ }\href@noop {} {\bibfield
  {journal} {\bibinfo  {journal} {Phys. Rev. Lett.}\ }\textbf {\bibinfo
  {volume} {92}},\ \bibinfo {pages} {017901} (\bibinfo {year}
  {2004})}\BibitemShut {NoStop}%
\bibitem [{\citenamefont {Li}\ \emph {et~al.}(2008)\citenamefont {Li},
  \citenamefont {Deng},\ and\ \citenamefont {Zhou}}]{Li08}%
  \BibitemOpen
  \bibfield  {author} {\bibinfo {author} {\bibfnamefont {X.-H.}\ \bibnamefont
  {Li}}, \bibinfo {author} {\bibfnamefont {F.-G.}\ \bibnamefont {Deng}}, \ and\
  \bibinfo {author} {\bibfnamefont {H.-Y.}\ \bibnamefont {Zhou}},\ }\href@noop
  {} {\bibfield  {journal} {\bibinfo  {journal} {Phys. Rev. A}\ }\textbf
  {\bibinfo {volume} {78}},\ \bibinfo {pages} {022321} (\bibinfo {year}
  {2008})}\BibitemShut {NoStop}%
\bibitem [{\citenamefont {Wang}(2004{\natexlab{a}})}]{Wang04a}%
  \BibitemOpen
  \bibfield  {author} {\bibinfo {author} {\bibfnamefont {X.-B.}\ \bibnamefont
  {Wang}},\ }\href@noop {} {\bibfield  {journal} {\bibinfo  {journal} {Phys.
  Rev. Lett.}\ }\textbf {\bibinfo {volume} {92}},\ \bibinfo {pages} {077902}
  (\bibinfo {year} {2004}{\natexlab{a}})}\BibitemShut {NoStop}%
\bibitem [{\citenamefont {Wang}(2004{\natexlab{b}})}]{Wang04b}%
  \BibitemOpen
  \bibfield  {author} {\bibinfo {author} {\bibfnamefont {X.-B.}\ \bibnamefont
  {Wang}},\ }\href@noop {} {\bibfield  {journal} {\bibinfo  {journal} {Phys.
  Rev. A}\ }\textbf {\bibinfo {volume} {69}},\ \bibinfo {pages} {022320}
  (\bibinfo {year} {2004}{\natexlab{b}})}\BibitemShut {NoStop}%
\bibitem [{\citenamefont {Kalamidas}(2005)}]{Kalamidas05}%
  \BibitemOpen
  \bibfield  {author} {\bibinfo {author} {\bibfnamefont {D.}~\bibnamefont
  {Kalamidas}},\ }\href@noop {} {\bibfield  {journal} {\bibinfo  {journal}
  {Phys. Lett. A}\ }\textbf {\bibinfo {volume} {343}},\ \bibinfo {pages} {331}
  (\bibinfo {year} {2005})}\BibitemShut {NoStop}%
\bibitem [{\citenamefont {Chen}\ \emph {et~al.}(2006)\citenamefont {Chen},
  \citenamefont {Zhang}, \citenamefont {Zhao}, \citenamefont {Zhou},\ and\
  \citenamefont {Pan}}]{Chen06}%
  \BibitemOpen
  \bibfield  {author} {\bibinfo {author} {\bibfnamefont {Y.-A.}\ \bibnamefont
  {Chen}}, \bibinfo {author} {\bibfnamefont {A.-N.}\ \bibnamefont {Zhang}},
  \bibinfo {author} {\bibfnamefont {Z.}~\bibnamefont {Zhao}}, \bibinfo {author}
  {\bibfnamefont {X.-Q.}\ \bibnamefont {Zhou}}, \ and\ \bibinfo {author}
  {\bibfnamefont {J.-W.}\ \bibnamefont {Pan}},\ }\href@noop {} {\bibfield
  {journal} {\bibinfo  {journal} {Phys. Rev. Lett.}\ }\textbf {\bibinfo
  {volume} {96}},\ \bibinfo {pages} {220504} (\bibinfo {year}
  {2006})}\BibitemShut {NoStop}%
\bibitem [{\citenamefont {Cai}\ and\ \citenamefont {Li}(2004)}]{Cai04}%
  \BibitemOpen
  \bibfield  {author} {\bibinfo {author} {\bibfnamefont {Q.-Y.}\ \bibnamefont
  {Cai}}\ and\ \bibinfo {author} {\bibfnamefont {B.-W.}\ \bibnamefont {Li}},\
  }\href@noop {} {\bibfield  {journal} {\bibinfo  {journal} {Phys. Rev. A}\
  }\textbf {\bibinfo {volume} {69}},\ \bibinfo {pages} {054301} (\bibinfo
  {year} {2004})}\BibitemShut {NoStop}%
\bibitem [{\citenamefont {Wang}\ \emph {et~al.}(2005)\citenamefont {Wang},
  \citenamefont {Deng}, \citenamefont {Li}, \citenamefont {Liu},\ and\
  \citenamefont {Long}}]{Wang05b}%
  \BibitemOpen
  \bibfield  {author} {\bibinfo {author} {\bibfnamefont {C.}~\bibnamefont
  {Wang}}, \bibinfo {author} {\bibfnamefont {F.-G.}\ \bibnamefont {Deng}},
  \bibinfo {author} {\bibfnamefont {Y.-S.}\ \bibnamefont {Li}}, \bibinfo
  {author} {\bibfnamefont {X.-S.}\ \bibnamefont {Liu}}, \ and\ \bibinfo
  {author} {\bibfnamefont {G.~L.}\ \bibnamefont {Long}},\ }\href@noop {}
  {\bibfield  {journal} {\bibinfo  {journal} {\emph{ibid.}}\ }\textbf {\bibinfo
  {volume} {71}},\ \bibinfo {pages} {044305} (\bibinfo {year}
  {2005})}\BibitemShut {NoStop}%
\bibitem [{\citenamefont {Li}\ \emph {et~al.}(2009)\citenamefont {Li},
  \citenamefont {Zhao}, \citenamefont {Sheng}, \citenamefont {Deng},\ and\
  \citenamefont {Zhou}}]{Li09}%
  \BibitemOpen
  \bibfield  {author} {\bibinfo {author} {\bibfnamefont {X.-H.}\ \bibnamefont
  {Li}}, \bibinfo {author} {\bibfnamefont {B.-K.}\ \bibnamefont {Zhao}},
  \bibinfo {author} {\bibfnamefont {Y.-B.}\ \bibnamefont {Sheng}}, \bibinfo
  {author} {\bibfnamefont {F.-G.}\ \bibnamefont {Deng}}, \ and\ \bibinfo
  {author} {\bibfnamefont {H.-Y.}\ \bibnamefont {Zhou}},\ }\href@noop {}
  {\bibfield  {journal} {\bibinfo  {journal} {Int. J. Quant. Inf.}\ }\textbf
  {\bibinfo {volume} {7}},\ \bibinfo {pages} {1479} (\bibinfo {year}
  {2009})}\BibitemShut {NoStop}%
\bibitem [{\citenamefont {Yang}\ and\ \citenamefont {Hwang}(2013)}]{Yang13}%
  \BibitemOpen
  \bibfield  {author} {\bibinfo {author} {\bibfnamefont {C.-W.}\ \bibnamefont
  {Yang}}\ and\ \bibinfo {author} {\bibfnamefont {T.}~\bibnamefont {Hwang}},\
  }\href@noop {} {\bibfield  {journal} {\bibinfo  {journal} {Quantum Inf.
  Process.}\ }\textbf {\bibinfo {volume} {12}},\ \bibinfo {pages} {3207}
  (\bibinfo {year} {2013})}\BibitemShut {NoStop}%
\bibitem [{\citenamefont {Eraerds}\ \emph {et~al.}(2010)\citenamefont
  {Eraerds}, \citenamefont {Walenta}, \citenamefont {Legr\'{e}}, \citenamefont
  {Gisin},\ and\ \citenamefont {Zbinden}}]{Eraerds10}%
  \BibitemOpen
  \bibfield  {author} {\bibinfo {author} {\bibfnamefont {P.}~\bibnamefont
  {Eraerds}}, \bibinfo {author} {\bibfnamefont {N.}~\bibnamefont {Walenta}},
  \bibinfo {author} {\bibfnamefont {M.}~\bibnamefont {Legr\'{e}}}, \bibinfo
  {author} {\bibfnamefont {N.}~\bibnamefont {Gisin}}, \ and\ \bibinfo {author}
  {\bibfnamefont {H.}~\bibnamefont {Zbinden}},\ }\href@noop {} {\bibfield
  {journal} {\bibinfo  {journal} {New J. Phys.}\ }\textbf {\bibinfo {volume}
  {12}},\ \bibinfo {pages} {063027} (\bibinfo {year} {2010})}\BibitemShut
  {NoStop}%
\bibitem [{\citenamefont {Qi}\ \emph {et~al.}(2010)\citenamefont {Qi},
  \citenamefont {Zhu}, \citenamefont {Qian},\ and\ \citenamefont {Lo}}]{Qi10}%
  \BibitemOpen
  \bibfield  {author} {\bibinfo {author} {\bibfnamefont {B.}~\bibnamefont
  {Qi}}, \bibinfo {author} {\bibfnamefont {W.}~\bibnamefont {Zhu}}, \bibinfo
  {author} {\bibfnamefont {L.}~\bibnamefont {Qian}}, \ and\ \bibinfo {author}
  {\bibfnamefont {H.-K.}\ \bibnamefont {Lo}},\ }\href@noop {} {\bibfield
  {journal} {\bibinfo  {journal} {New J. Phys.}\ }\textbf {\bibinfo {volume}
  {12}},\ \bibinfo {pages} {103042} (\bibinfo {year} {2010})}\BibitemShut
  {NoStop}%
\bibitem [{\citenamefont {Scarani}\ \emph {et~al.}(2009)\citenamefont
  {Scarani}, \citenamefont {Bechmann-Pasquinucci}, \citenamefont {Cerf},
  \citenamefont {Du\v{s}ek}, \citenamefont {L\"{u}tkenhaus},\ and\
  \citenamefont {Peev}}]{Scarani09}%
  \BibitemOpen
  \bibfield  {author} {\bibinfo {author} {\bibfnamefont {V.}~\bibnamefont
  {Scarani}}, \bibinfo {author} {\bibfnamefont {H.}~\bibnamefont
  {Bechmann-Pasquinucci}}, \bibinfo {author} {\bibfnamefont {N.~J.}\
  \bibnamefont {Cerf}}, \bibinfo {author} {\bibfnamefont {M.}~\bibnamefont
  {Du\v{s}ek}}, \bibinfo {author} {\bibfnamefont {N.}~\bibnamefont
  {L\"{u}tkenhaus}}, \ and\ \bibinfo {author} {\bibfnamefont {M.}~\bibnamefont
  {Peev}},\ }\href@noop {} {\bibfield  {journal} {\bibinfo  {journal} {Rev.
  Mod. Phys.}\ }\textbf {\bibinfo {volume} {81}},\ \bibinfo {pages} {1301}
  (\bibinfo {year} {2009})}\BibitemShut {NoStop}%
\bibitem [{\citenamefont {Csisz{\'a}r}\ and\ \citenamefont
  {K{\"o}rner}(1978)}]{Csiszar1978}%
  \BibitemOpen
  \bibfield  {author} {\bibinfo {author} {\bibfnamefont {I.}~\bibnamefont
  {Csisz{\'a}r}}\ and\ \bibinfo {author} {\bibfnamefont {J.}~\bibnamefont
  {K{\"o}rner}},\ }\href@noop {} {\bibfield  {journal} {\bibinfo  {journal}
  {IEEE Trans. Inf. Theory}\ }\textbf {\bibinfo {volume} {24}},\ \bibinfo
  {pages} {339} (\bibinfo {year} {1978})}\BibitemShut {NoStop}%
\bibitem [{\citenamefont {Devetak}\ and\ \citenamefont
  {Winter}(2005)}]{Devetak2005}%
  \BibitemOpen
  \bibfield  {author} {\bibinfo {author} {\bibfnamefont {I.}~\bibnamefont
  {Devetak}}\ and\ \bibinfo {author} {\bibfnamefont {A.}~\bibnamefont
  {Winter}},\ }\href@noop {} {\bibfield  {journal} {\bibinfo  {journal} {Proc.
  R. Soc. London, Ser. A}\ }\textbf {\bibinfo {volume} {461}},\ \bibinfo
  {pages} {207} (\bibinfo {year} {2005})}\BibitemShut {NoStop}%
\bibitem [{\citenamefont {H.-K}\ \emph {et~al.}(2005)\citenamefont {H.-K},
  \citenamefont {Lo}, \citenamefont {Chau},\ and\ \citenamefont
  {Ardehali}}]{Lo05b}%
  \BibitemOpen
  \bibfield  {author} {\bibinfo {author} {\bibnamefont {H.-K}}, \bibinfo
  {author} {\bibnamefont {Lo}}, \bibinfo {author} {\bibfnamefont {H.~F.}\
  \bibnamefont {Chau}}, \ and\ \bibinfo {author} {\bibfnamefont
  {M.}~\bibnamefont {Ardehali}},\ }\href@noop {} {\bibfield  {journal}
  {\bibinfo  {journal} {J. Cryptology}\ }\textbf {\bibinfo {volume} {18}},\
  \bibinfo {pages} {133} (\bibinfo {year} {2005})}\BibitemShut {NoStop}%
\bibitem [{\citenamefont {Wolf}\ \emph {et~al.}(2006)\citenamefont {Wolf},
  \citenamefont {Giedke},\ and\ \citenamefont {Cirac}}]{Wolf2006}%
  \BibitemOpen
  \bibfield  {author} {\bibinfo {author} {\bibfnamefont {M.~M.}\ \bibnamefont
  {Wolf}}, \bibinfo {author} {\bibfnamefont {G.}~\bibnamefont {Giedke}}, \ and\
  \bibinfo {author} {\bibfnamefont {J.~I.}\ \bibnamefont {Cirac}},\ }\href@noop
  {} {\bibfield  {journal} {\bibinfo  {journal} {Phys. Rev. Lett.}\ }\textbf
  {\bibinfo {volume} {96}},\ \bibinfo {pages} {080502} (\bibinfo {year}
  {2006})}\BibitemShut {NoStop}%
\bibitem [{\citenamefont {Renner}\ and\ \citenamefont
  {Cirac}(2009)}]{Renner2009}%
  \BibitemOpen
  \bibfield  {author} {\bibinfo {author} {\bibfnamefont {R.}~\bibnamefont
  {Renner}}\ and\ \bibinfo {author} {\bibfnamefont {J.~I.}\ \bibnamefont
  {Cirac}},\ }\href@noop {} {\bibfield  {journal} {\bibinfo  {journal} {Phys.
  Rev. Lett.}\ }\textbf {\bibinfo {volume} {102}},\ \bibinfo {pages} {110504}
  (\bibinfo {year} {2009})}\BibitemShut {NoStop}%
\bibitem [{\citenamefont {Holevo}\ and\ \citenamefont
  {Werner}(2001)}]{Holevo2001}%
  \BibitemOpen
  \bibfield  {author} {\bibinfo {author} {\bibfnamefont {A.~S.}\ \bibnamefont
  {Holevo}}\ and\ \bibinfo {author} {\bibfnamefont {R.~F.}\ \bibnamefont
  {Werner}},\ }\href@noop {} {\bibfield  {journal} {\bibinfo  {journal} {Phys.
  Rev. A}\ }\textbf {\bibinfo {volume} {63}},\ \bibinfo {pages} {032312}
  (\bibinfo {year} {2001})}\BibitemShut {NoStop}%
\bibitem [{\citenamefont {Serafini}\ \emph {et~al.}(2005)\citenamefont
  {Serafini}, \citenamefont {Paris}, \citenamefont {Illuminati},\ and\
  \citenamefont {De~Siena}}]{Serafini2005}%
  \BibitemOpen
  \bibfield  {author} {\bibinfo {author} {\bibfnamefont {A.}~\bibnamefont
  {Serafini}}, \bibinfo {author} {\bibfnamefont {M.}~\bibnamefont {Paris}},
  \bibinfo {author} {\bibfnamefont {F.}~\bibnamefont {Illuminati}}, \ and\
  \bibinfo {author} {\bibfnamefont {S.}~\bibnamefont {De~Siena}},\ }\href@noop
  {} {\bibfield  {journal} {\bibinfo  {journal} {J. Opt. B: Quantum S. O.}\
  }\textbf {\bibinfo {volume} {7}},\ \bibinfo {pages} {R19} (\bibinfo {year}
  {2005})}\BibitemShut {NoStop}%
\bibitem [{\citenamefont {Bruss}(1998)}]{Bruss98}%
  \BibitemOpen
  \bibfield  {author} {\bibinfo {author} {\bibfnamefont {D.}~\bibnamefont
  {Bruss}},\ }\href@noop {} {\bibfield  {journal} {\bibinfo  {journal} {Phys.
  Rev. Lett.}\ }\textbf {\bibinfo {volume} {81}},\ \bibinfo {pages} {3018}
  (\bibinfo {year} {1998})}\BibitemShut {NoStop}%
\bibitem [{\citenamefont {Hwang}(2003)}]{Hwang03}%
  \BibitemOpen
  \bibfield  {author} {\bibinfo {author} {\bibfnamefont {W.-Y.}\ \bibnamefont
  {Hwang}},\ }\href@noop {} {\bibfield  {journal} {\bibinfo  {journal} {Phys.
  Rev. Lett.}\ }\textbf {\bibinfo {volume} {91}},\ \bibinfo {pages} {057901}
  (\bibinfo {year} {2003})}\BibitemShut {NoStop}%
\bibitem [{\citenamefont {Wang}(2005)}]{Wang05a}%
  \BibitemOpen
  \bibfield  {author} {\bibinfo {author} {\bibfnamefont {X.-B.}\ \bibnamefont
  {Wang}},\ }\href@noop {} {\bibfield  {journal} {\bibinfo  {journal}
  {\emph{ibid.}}\ }\textbf {\bibinfo {volume} {94}},\ \bibinfo {pages} {230503}
  (\bibinfo {year} {2005})}\BibitemShut {NoStop}%
\bibitem [{\citenamefont {Lo}\ \emph {et~al.}(2005)\citenamefont {Lo},
  \citenamefont {Ma},\ and\ \citenamefont {Chen}}]{Lo05}%
  \BibitemOpen
  \bibfield  {author} {\bibinfo {author} {\bibfnamefont {H.-K.}\ \bibnamefont
  {Lo}}, \bibinfo {author} {\bibfnamefont {X.}~\bibnamefont {Ma}}, \ and\
  \bibinfo {author} {\bibfnamefont {K.}~\bibnamefont {Chen}},\ }\href@noop {}
  {\bibfield  {journal} {\bibinfo  {journal} {\emph{ibid}.}\ }\textbf {\bibinfo
  {volume} {94}},\ \bibinfo {pages} {230504} (\bibinfo {year}
  {2005})}\BibitemShut {NoStop}%
\bibitem [{\citenamefont {Fasel}\ \emph {et~al.}(2004)\citenamefont {Fasel},
  \citenamefont {Alibart}, \citenamefont {Tanzilli}, \citenamefont {Baldi},
  \citenamefont {Beveratos}, \citenamefont {Gisin},\ and\ \citenamefont
  {Zbinden}}]{Fasel04}%
  \BibitemOpen
  \bibfield  {author} {\bibinfo {author} {\bibfnamefont {S.}~\bibnamefont
  {Fasel}}, \bibinfo {author} {\bibfnamefont {O.}~\bibnamefont {Alibart}},
  \bibinfo {author} {\bibfnamefont {S.}~\bibnamefont {Tanzilli}}, \bibinfo
  {author} {\bibfnamefont {P.}~\bibnamefont {Baldi}}, \bibinfo {author}
  {\bibfnamefont {A.}~\bibnamefont {Beveratos}}, \bibinfo {author}
  {\bibfnamefont {N.}~\bibnamefont {Gisin}}, \ and\ \bibinfo {author}
  {\bibfnamefont {H.}~\bibnamefont {Zbinden}},\ }\href@noop {} {\bibfield
  {journal} {\bibinfo  {journal} {New J. Phys.}\ }\textbf {\bibinfo {volume}
  {6}},\ \bibinfo {pages} {163} (\bibinfo {year} {2004})}\BibitemShut {NoStop}%
\bibitem [{\citenamefont {Keller}\ \emph {et~al.}(2004)\citenamefont {Keller},
  \citenamefont {Lange}, \citenamefont {Hayasaka}, \citenamefont {Lange},\ and\
  \citenamefont {Walther}}]{Keller04}%
  \BibitemOpen
  \bibfield  {author} {\bibinfo {author} {\bibfnamefont {M.}~\bibnamefont
  {Keller}}, \bibinfo {author} {\bibfnamefont {B.}~\bibnamefont {Lange}},
  \bibinfo {author} {\bibfnamefont {K.}~\bibnamefont {Hayasaka}}, \bibinfo
  {author} {\bibfnamefont {W.}~\bibnamefont {Lange}}, \ and\ \bibinfo {author}
  {\bibfnamefont {H.}~\bibnamefont {Walther}},\ }\href@noop {} {\bibfield
  {journal} {\bibinfo  {journal} {Nature (London)}\ }\textbf {\bibinfo {volume}
  {431}},\ \bibinfo {pages} {1075} (\bibinfo {year} {2004})}\BibitemShut
  {NoStop}%
\bibitem [{\citenamefont {Brokmann}\ \emph {et~al.}(2004)\citenamefont
  {Brokmann}, \citenamefont {Giacobino}, \citenamefont {Dahan},\ and\
  \citenamefont {Hermier}}]{Brokmann04}%
  \BibitemOpen
  \bibfield  {author} {\bibinfo {author} {\bibfnamefont {X.}~\bibnamefont
  {Brokmann}}, \bibinfo {author} {\bibfnamefont {E.}~\bibnamefont {Giacobino}},
  \bibinfo {author} {\bibfnamefont {M.}~\bibnamefont {Dahan}}, \ and\ \bibinfo
  {author} {\bibfnamefont {J.~P.}\ \bibnamefont {Hermier}},\ }\href@noop {}
  {\bibfield  {journal} {\bibinfo  {journal} {Appl. Phys. Lett.}\ }\textbf
  {\bibinfo {volume} {85}},\ \bibinfo {pages} {712} (\bibinfo {year}
  {2004})}\BibitemShut {NoStop}%
\bibitem [{\citenamefont {Laurat}\ \emph {et~al.}(2006)\citenamefont {Laurat},
  \citenamefont {de~Riedmatten}, \citenamefont {Felinto}, \citenamefont {Chou},
  \citenamefont {Schomburg},\ and\ \citenamefont {Kimble}}]{Laurat06}%
  \BibitemOpen
  \bibfield  {author} {\bibinfo {author} {\bibfnamefont {J.}~\bibnamefont
  {Laurat}}, \bibinfo {author} {\bibfnamefont {H.}~\bibnamefont
  {de~Riedmatten}}, \bibinfo {author} {\bibfnamefont {D.}~\bibnamefont
  {Felinto}}, \bibinfo {author} {\bibfnamefont {C.-W.}\ \bibnamefont {Chou}},
  \bibinfo {author} {\bibfnamefont {E.~W.}\ \bibnamefont {Schomburg}}, \ and\
  \bibinfo {author} {\bibfnamefont {H.~J.}\ \bibnamefont {Kimble}},\
  }\href@noop {} {\bibfield  {journal} {\bibinfo  {journal} {Opt. Express}\
  }\textbf {\bibinfo {volume} {14}},\ \bibinfo {pages} {6912} (\bibinfo {year}
  {2006})}\BibitemShut {NoStop}%
\bibitem [{\citenamefont {Pisanello}\ \emph {et~al.}(2010)\citenamefont
  {Pisanello}, \citenamefont {Martiradonna}, \citenamefont {Lem\'{e}nager},
  \citenamefont {Spinicelli}, \citenamefont {Fiore}, \citenamefont {Manna},
  \citenamefont {Hermier}, \citenamefont {Cingolani}, \citenamefont
  {Giacobino}, \citenamefont {{De Vittorio}},\ and\ \citenamefont
  {Bramati}}]{Pisanello10}%
  \BibitemOpen
  \bibfield  {author} {\bibinfo {author} {\bibfnamefont {F.}~\bibnamefont
  {Pisanello}}, \bibinfo {author} {\bibfnamefont {L.}~\bibnamefont
  {Martiradonna}}, \bibinfo {author} {\bibfnamefont {G.}~\bibnamefont
  {Lem\'{e}nager}}, \bibinfo {author} {\bibfnamefont {P.}~\bibnamefont
  {Spinicelli}}, \bibinfo {author} {\bibfnamefont {A.}~\bibnamefont {Fiore}},
  \bibinfo {author} {\bibfnamefont {L.}~\bibnamefont {Manna}}, \bibinfo
  {author} {\bibfnamefont {J.-P.}\ \bibnamefont {Hermier}}, \bibinfo {author}
  {\bibfnamefont {R.}~\bibnamefont {Cingolani}}, \bibinfo {author}
  {\bibfnamefont {E.}~\bibnamefont {Giacobino}}, \bibinfo {author}
  {\bibfnamefont {M.}~\bibnamefont {{De Vittorio}}}, \ and\ \bibinfo {author}
  {\bibfnamefont {A.}~\bibnamefont {Bramati}},\ }\href@noop {} {\bibfield
  {journal} {\bibinfo  {journal} {Appl. Phys. Lett.}\ }\textbf {\bibinfo
  {volume} {96}},\ \bibinfo {pages} {033101} (\bibinfo {year}
  {2010})}\BibitemShut {NoStop}%
\bibitem [{\citenamefont {M\"{u}cke}\ \emph {et~al.}(2013)\citenamefont
  {M\"{u}cke}, \citenamefont {Bochmann}, \citenamefont {Hahn}, \citenamefont
  {Neuzner}, \citenamefont {N\"{o}lleke}, \citenamefont {Reiserer},
  \citenamefont {Rempe},\ and\ \citenamefont {Ritter}}]{Mucke13}%
  \BibitemOpen
  \bibfield  {author} {\bibinfo {author} {\bibfnamefont {M.}~\bibnamefont
  {M\"{u}cke}}, \bibinfo {author} {\bibfnamefont {J.}~\bibnamefont {Bochmann}},
  \bibinfo {author} {\bibfnamefont {C.}~\bibnamefont {Hahn}}, \bibinfo {author}
  {\bibfnamefont {A.}~\bibnamefont {Neuzner}}, \bibinfo {author} {\bibfnamefont
  {C.}~\bibnamefont {N\"{o}lleke}}, \bibinfo {author} {\bibfnamefont
  {A.}~\bibnamefont {Reiserer}}, \bibinfo {author} {\bibfnamefont
  {G.}~\bibnamefont {Rempe}}, \ and\ \bibinfo {author} {\bibfnamefont
  {S.}~\bibnamefont {Ritter}},\ }\href@noop {} {\bibfield  {journal} {\bibinfo
  {journal} {Phys. Rev. A}\ }\textbf {\bibinfo {volume} {87}},\ \bibinfo
  {pages} {063805} (\bibinfo {year} {2013})}\BibitemShut {NoStop}%
\bibitem [{\citenamefont {Claudon}\ \emph {et~al.}(2010)\citenamefont
  {Claudon}, \citenamefont {Bleuse}, \citenamefont {Malik}, \citenamefont
  {Bazin}, \citenamefont {Jaffrennou}, \citenamefont {Gregersen}, \citenamefont
  {Sauvan}, \citenamefont {Lalanne},\ and\ \citenamefont
  {G\'{e}rard}}]{Claudon10}%
  \BibitemOpen
  \bibfield  {author} {\bibinfo {author} {\bibfnamefont {J.}~\bibnamefont
  {Claudon}}, \bibinfo {author} {\bibfnamefont {J.}~\bibnamefont {Bleuse}},
  \bibinfo {author} {\bibfnamefont {N.~S.}\ \bibnamefont {Malik}}, \bibinfo
  {author} {\bibfnamefont {M.}~\bibnamefont {Bazin}}, \bibinfo {author}
  {\bibfnamefont {P.}~\bibnamefont {Jaffrennou}}, \bibinfo {author}
  {\bibfnamefont {N.}~\bibnamefont {Gregersen}}, \bibinfo {author}
  {\bibfnamefont {C.}~\bibnamefont {Sauvan}}, \bibinfo {author} {\bibfnamefont
  {P.}~\bibnamefont {Lalanne}}, \ and\ \bibinfo {author} {\bibfnamefont
  {J.-M.}\ \bibnamefont {G\'{e}rard}},\ }\href@noop {} {\bibfield  {journal}
  {\bibinfo  {journal} {Nat. Photon.}\ }\textbf {\bibinfo {volume} {4}},\
  \bibinfo {pages} {174} (\bibinfo {year} {2010})}\BibitemShut {NoStop}%
\bibitem [{\citenamefont {Eisaman}\ \emph {et~al.}(2011)\citenamefont
  {Eisaman}, \citenamefont {Fan}, \citenamefont {Migdall},\ and\ \citenamefont
  {Polyakov}}]{Eisaman11}%
  \BibitemOpen
  \bibfield  {author} {\bibinfo {author} {\bibfnamefont {M.~D.}\ \bibnamefont
  {Eisaman}}, \bibinfo {author} {\bibfnamefont {J.}~\bibnamefont {Fan}},
  \bibinfo {author} {\bibfnamefont {A.}~\bibnamefont {Migdall}}, \ and\
  \bibinfo {author} {\bibfnamefont {S.~V.}\ \bibnamefont {Polyakov}},\
  }\href@noop {} {\bibfield  {journal} {\bibinfo  {journal} {Rev. Sci
  Instrum.}\ }\textbf {\bibinfo {volume} {82}},\ \bibinfo {pages} {071101}
  (\bibinfo {year} {2011})}\BibitemShut {NoStop}%
\bibitem [{\citenamefont {Bulgarini}\ \emph {et~al.}(2012)\citenamefont
  {Bulgarini}, \citenamefont {Reimer}, \citenamefont {Zehender}, \citenamefont
  {Hocevar}, \citenamefont {Bakkers}, \citenamefont {Kouwenhoven},\ and\
  \citenamefont {Zwiller}}]{Bulgarini12}%
  \BibitemOpen
  \bibfield  {author} {\bibinfo {author} {\bibfnamefont {G.}~\bibnamefont
  {Bulgarini}}, \bibinfo {author} {\bibfnamefont {M.~E.}\ \bibnamefont
  {Reimer}}, \bibinfo {author} {\bibfnamefont {T.}~\bibnamefont {Zehender}},
  \bibinfo {author} {\bibfnamefont {M.}~\bibnamefont {Hocevar}}, \bibinfo
  {author} {\bibfnamefont {E.~P. A.~M.}\ \bibnamefont {Bakkers}}, \bibinfo
  {author} {\bibfnamefont {L.~P.}\ \bibnamefont {Kouwenhoven}}, \ and\ \bibinfo
  {author} {\bibfnamefont {V.}~\bibnamefont {Zwiller}},\ }\href@noop {}
  {\bibfield  {journal} {\bibinfo  {journal} {Appl. Phys. Lett.}\ }\textbf
  {\bibinfo {volume} {100}},\ \bibinfo {pages} {121106} (\bibinfo {year}
  {2012})}\BibitemShut {NoStop}%
\bibitem [{\citenamefont {Gazzano}\ \emph {et~al.}(2012)\citenamefont
  {Gazzano}, \citenamefont {{Michaelis de Vasconcellos}}, \citenamefont
  {Arnold}, \citenamefont {Nowak}, \citenamefont {Galopin}, \citenamefont
  {Sagnes}, \citenamefont {Lanco}, \citenamefont {Lema\^{i}tre},\ and\
  \citenamefont {Senellart}}]{Gazzano12}%
  \BibitemOpen
  \bibfield  {author} {\bibinfo {author} {\bibfnamefont {O.}~\bibnamefont
  {Gazzano}}, \bibinfo {author} {\bibfnamefont {S.}~\bibnamefont {{Michaelis de
  Vasconcellos}}}, \bibinfo {author} {\bibfnamefont {C.}~\bibnamefont
  {Arnold}}, \bibinfo {author} {\bibfnamefont {A.}~\bibnamefont {Nowak}},
  \bibinfo {author} {\bibfnamefont {E.}~\bibnamefont {Galopin}}, \bibinfo
  {author} {\bibfnamefont {I.}~\bibnamefont {Sagnes}}, \bibinfo {author}
  {\bibfnamefont {L.}~\bibnamefont {Lanco}}, \bibinfo {author} {\bibfnamefont
  {A.}~\bibnamefont {Lema\^{i}tre}}, \ and\ \bibinfo {author} {\bibfnamefont
  {P.}~\bibnamefont {Senellart}},\ }\href@noop {} {\bibfield  {journal}
  {\bibinfo  {journal} {Nat. Commun.}\ }\textbf {\bibinfo {volume} {4}},\
  \bibinfo {pages} {1425} (\bibinfo {year} {2012})}\BibitemShut {NoStop}%
\bibitem [{\citenamefont {Somaschi}\ \emph {et~al.}(2016)\citenamefont
  {Somaschi}, \citenamefont {Giesz}, \citenamefont {{De Santis}}, \citenamefont
  {Loredo}, \citenamefont {Almeida}, \citenamefont {Hornecker}, \citenamefont
  {Portalupi}, \citenamefont {Grange}, \citenamefont {Ant\'{o}n}, \citenamefont
  {Demory}, \citenamefont {G\'{o}mez}, \citenamefont {Sagnes}, \citenamefont
  {Lanzillotti-Kimura}, \citenamefont {Lema\'{i}tre}, \citenamefont {Auffeves},
  \citenamefont {White}, \citenamefont {Lanco},\ and\ \citenamefont
  {Sennelart}}]{Somaschi16}%
  \BibitemOpen
  \bibfield  {author} {\bibinfo {author} {\bibfnamefont {N.}~\bibnamefont
  {Somaschi}}, \bibinfo {author} {\bibfnamefont {V.}~\bibnamefont {Giesz}},
  \bibinfo {author} {\bibfnamefont {L.}~\bibnamefont {{De Santis}}}, \bibinfo
  {author} {\bibfnamefont {J.~C.}\ \bibnamefont {Loredo}}, \bibinfo {author}
  {\bibfnamefont {M.~P.}\ \bibnamefont {Almeida}}, \bibinfo {author}
  {\bibfnamefont {G.}~\bibnamefont {Hornecker}}, \bibinfo {author}
  {\bibfnamefont {S.~L.}\ \bibnamefont {Portalupi}}, \bibinfo {author}
  {\bibfnamefont {T.}~\bibnamefont {Grange}}, \bibinfo {author} {\bibfnamefont
  {C.}~\bibnamefont {Ant\'{o}n}}, \bibinfo {author} {\bibfnamefont
  {J.}~\bibnamefont {Demory}}, \bibinfo {author} {\bibfnamefont
  {C.}~\bibnamefont {G\'{o}mez}}, \bibinfo {author} {\bibfnamefont
  {I.}~\bibnamefont {Sagnes}}, \bibinfo {author} {\bibfnamefont {N.~D.}\
  \bibnamefont {Lanzillotti-Kimura}}, \bibinfo {author} {\bibfnamefont
  {A.}~\bibnamefont {Lema\'{i}tre}}, \bibinfo {author} {\bibfnamefont
  {A.}~\bibnamefont {Auffeves}}, \bibinfo {author} {\bibfnamefont {A.~G.}\
  \bibnamefont {White}}, \bibinfo {author} {\bibfnamefont {L.}~\bibnamefont
  {Lanco}}, \ and\ \bibinfo {author} {\bibfnamefont {P.}~\bibnamefont
  {Sennelart}},\ }\href@noop {} {\bibfield  {journal} {\bibinfo  {journal}
  {Nat. Photon.}\ }\textbf {\bibinfo {volume} {10}},\ \bibinfo {pages} {340}
  (\bibinfo {year} {2016})}\BibitemShut {NoStop}%
\bibitem [{\citenamefont {Pomarico}\ \emph {et~al.}(2012)\citenamefont
  {Pomarico}, \citenamefont {Sanguinetti}, \citenamefont {Guerreiro},
  \citenamefont {Thew},\ and\ \citenamefont {Zbinden}}]{Pomarico12}%
  \BibitemOpen
  \bibfield  {author} {\bibinfo {author} {\bibfnamefont {E.}~\bibnamefont
  {Pomarico}}, \bibinfo {author} {\bibfnamefont {B.}~\bibnamefont
  {Sanguinetti}}, \bibinfo {author} {\bibfnamefont {T.}~\bibnamefont
  {Guerreiro}}, \bibinfo {author} {\bibfnamefont {R.}~\bibnamefont {Thew}}, \
  and\ \bibinfo {author} {\bibfnamefont {H.}~\bibnamefont {Zbinden}},\
  }\href@noop {} {\bibfield  {journal} {\bibinfo  {journal} {Opt. Express}\
  }\textbf {\bibinfo {volume} {20}},\ \bibinfo {pages} {23846} (\bibinfo {year}
  {2012})}\BibitemShut {NoStop}%
\bibitem [{\citenamefont {{Da Cunha Pereira}}\ \emph
  {et~al.}(2013)\citenamefont {{Da Cunha Pereira}}, \citenamefont {Becerra},
  \citenamefont {Glebov}, \citenamefont {Fan}, \citenamefont {Nam},\ and\
  \citenamefont {Migdall}}]{Pereira13}%
  \BibitemOpen
  \bibfield  {author} {\bibinfo {author} {\bibfnamefont {M.}~\bibnamefont {{Da
  Cunha Pereira}}}, \bibinfo {author} {\bibfnamefont {F.~E.}\ \bibnamefont
  {Becerra}}, \bibinfo {author} {\bibfnamefont {B.~L.}\ \bibnamefont {Glebov}},
  \bibinfo {author} {\bibfnamefont {J.}~\bibnamefont {Fan}}, \bibinfo {author}
  {\bibfnamefont {S.~W.}\ \bibnamefont {Nam}}, \ and\ \bibinfo {author}
  {\bibfnamefont {A.}~\bibnamefont {Migdall}},\ }\href@noop {} {\bibfield
  {journal} {\bibinfo  {journal} {Opt. Lett.}\ }\textbf {\bibinfo {volume}
  {38}},\ \bibinfo {pages} {1609} (\bibinfo {year} {2013})}\BibitemShut
  {NoStop}%
\bibitem [{\citenamefont {Ramelow}\ \emph {et~al.}(2013)\citenamefont
  {Ramelow}, \citenamefont {Mech}, \citenamefont {Giustina}, \citenamefont
  {Gr\"{o}blacher}, \citenamefont {Wieczorek}, \citenamefont {Beyer},
  \citenamefont {Lita}, \citenamefont {Calkins}, \citenamefont {Gerrits},
  \citenamefont {Nam}, \citenamefont {Zeilinger},\ and\ \citenamefont
  {Ursin}}]{Ramelow13}%
  \BibitemOpen
  \bibfield  {author} {\bibinfo {author} {\bibfnamefont {S.}~\bibnamefont
  {Ramelow}}, \bibinfo {author} {\bibfnamefont {A.}~\bibnamefont {Mech}},
  \bibinfo {author} {\bibfnamefont {M.}~\bibnamefont {Giustina}}, \bibinfo
  {author} {\bibfnamefont {S.}~\bibnamefont {Gr\"{o}blacher}}, \bibinfo
  {author} {\bibfnamefont {W.}~\bibnamefont {Wieczorek}}, \bibinfo {author}
  {\bibfnamefont {J.}~\bibnamefont {Beyer}}, \bibinfo {author} {\bibfnamefont
  {A.}~\bibnamefont {Lita}}, \bibinfo {author} {\bibfnamefont {B.}~\bibnamefont
  {Calkins}}, \bibinfo {author} {\bibfnamefont {T.}~\bibnamefont {Gerrits}},
  \bibinfo {author} {\bibfnamefont {S.~W.}\ \bibnamefont {Nam}}, \bibinfo
  {author} {\bibfnamefont {A.}~\bibnamefont {Zeilinger}}, \ and\ \bibinfo
  {author} {\bibfnamefont {R.}~\bibnamefont {Ursin}},\ }\href@noop {}
  {\bibfield  {journal} {\bibinfo  {journal} {Opt. Express}\ }\textbf {\bibinfo
  {volume} {21}},\ \bibinfo {pages} {6707} (\bibinfo {year}
  {2013})}\BibitemShut {NoStop}%
\bibitem [{\citenamefont {Usenko}\ and\ \citenamefont
  {Filip}(2011)}]{Usenko2011}%
  \BibitemOpen
  \bibfield  {author} {\bibinfo {author} {\bibfnamefont {V.~C.}\ \bibnamefont
  {Usenko}}\ and\ \bibinfo {author} {\bibfnamefont {R.}~\bibnamefont {Filip}},\
  }\href@noop {} {\bibfield  {journal} {\bibinfo  {journal} {New J. Phys.}\
  }\textbf {\bibinfo {volume} {13}},\ \bibinfo {pages} {113007} (\bibinfo
  {year} {2011})}\BibitemShut {NoStop}%
\bibitem [{\citenamefont {Zhang}\ \emph {et~al.}(2008)\citenamefont {Zhang},
  \citenamefont {Takesue}, \citenamefont {Nam}, \citenamefont {Langrock},
  \citenamefont {Xie}, \citenamefont {Baek}, \citenamefont {Fejer},\ and\
  \citenamefont {Yamamoto}}]{Zhang08}%
  \BibitemOpen
  \bibfield  {author} {\bibinfo {author} {\bibfnamefont {Q.}~\bibnamefont
  {Zhang}}, \bibinfo {author} {\bibfnamefont {H.}~\bibnamefont {Takesue}},
  \bibinfo {author} {\bibfnamefont {S.~W.}\ \bibnamefont {Nam}}, \bibinfo
  {author} {\bibfnamefont {C.}~\bibnamefont {Langrock}}, \bibinfo {author}
  {\bibfnamefont {X.}~\bibnamefont {Xie}}, \bibinfo {author} {\bibfnamefont
  {B.}~\bibnamefont {Baek}}, \bibinfo {author} {\bibfnamefont {M.~M.}\
  \bibnamefont {Fejer}}, \ and\ \bibinfo {author} {\bibfnamefont
  {Y.}~\bibnamefont {Yamamoto}},\ }\href@noop {} {\bibfield  {journal}
  {\bibinfo  {journal} {Opt. Express}\ }\textbf {\bibinfo {volume} {16}},\
  \bibinfo {pages} {5776} (\bibinfo {year} {2008})}\BibitemShut {NoStop}%
\bibitem [{\citenamefont {Walenta}\ \emph {et~al.}(2014)\citenamefont
  {Walenta}, \citenamefont {Burg}, \citenamefont {Caselunghe}, \citenamefont
  {Constantin}, \citenamefont {Gisin}, \citenamefont {Guinnard}, \citenamefont
  {Houlmann}, \citenamefont {Junod}, \citenamefont {Korzh}, \citenamefont
  {Kulesza}, \citenamefont {Legr\'{e}}, \citenamefont {Lim}, \citenamefont
  {Lunghi}, \citenamefont {Monat}, \citenamefont {Portmann}, \citenamefont
  {Soucarros}, \citenamefont {Thew}, \citenamefont {Trinkler}, \citenamefont
  {Trolliet}, \citenamefont {Vannel},\ and\ \citenamefont
  {Zbinden}}]{Walenta14}%
  \BibitemOpen
  \bibfield  {author} {\bibinfo {author} {\bibfnamefont {N.}~\bibnamefont
  {Walenta}}, \bibinfo {author} {\bibfnamefont {A.}~\bibnamefont {Burg}},
  \bibinfo {author} {\bibfnamefont {D.}~\bibnamefont {Caselunghe}}, \bibinfo
  {author} {\bibfnamefont {J.}~\bibnamefont {Constantin}}, \bibinfo {author}
  {\bibfnamefont {N.}~\bibnamefont {Gisin}}, \bibinfo {author} {\bibfnamefont
  {O.}~\bibnamefont {Guinnard}}, \bibinfo {author} {\bibfnamefont
  {R.}~\bibnamefont {Houlmann}}, \bibinfo {author} {\bibfnamefont
  {P.}~\bibnamefont {Junod}}, \bibinfo {author} {\bibfnamefont
  {B.}~\bibnamefont {Korzh}}, \bibinfo {author} {\bibfnamefont
  {N.}~\bibnamefont {Kulesza}}, \bibinfo {author} {\bibfnamefont
  {M.}~\bibnamefont {Legr\'{e}}}, \bibinfo {author} {\bibfnamefont {C.~W.}\
  \bibnamefont {Lim}}, \bibinfo {author} {\bibfnamefont {T.}~\bibnamefont
  {Lunghi}}, \bibinfo {author} {\bibfnamefont {L.}~\bibnamefont {Monat}},
  \bibinfo {author} {\bibfnamefont {C.}~\bibnamefont {Portmann}}, \bibinfo
  {author} {\bibfnamefont {M.}~\bibnamefont {Soucarros}}, \bibinfo {author}
  {\bibfnamefont {R.~T.}\ \bibnamefont {Thew}}, \bibinfo {author}
  {\bibfnamefont {P.}~\bibnamefont {Trinkler}}, \bibinfo {author}
  {\bibfnamefont {G.}~\bibnamefont {Trolliet}}, \bibinfo {author}
  {\bibfnamefont {F.}~\bibnamefont {Vannel}}, \ and\ \bibinfo {author}
  {\bibfnamefont {H.}~\bibnamefont {Zbinden}},\ }\href@noop {} {\bibfield
  {journal} {\bibinfo  {journal} {New J. Phys.}\ }\textbf {\bibinfo {volume}
  {16}},\ \bibinfo {pages} {013047} (\bibinfo {year} {2014})}\BibitemShut
  {NoStop}%
\bibitem [{\citenamefont {Valivarthi}\ \emph {et~al.}(2015)\citenamefont
  {Valivarthi}, \citenamefont {Lucio-Martinez}, \citenamefont {Chan},
  \citenamefont {Rubenok}, \citenamefont {John}, \citenamefont {Korchinski},
  \citenamefont {Duffin}, \citenamefont {Marsili}, \citenamefont {Verma},
  \citenamefont {Shaw}, \citenamefont {Stern}, \citenamefont {Nam},
  \citenamefont {Oblak}, \citenamefont {Zhou}, \citenamefont {Slater},\ and\
  \citenamefont {Tittel}}]{Valivarthi15}%
  \BibitemOpen
  \bibfield  {author} {\bibinfo {author} {\bibfnamefont {R.}~\bibnamefont
  {Valivarthi}}, \bibinfo {author} {\bibfnamefont {I.}~\bibnamefont
  {Lucio-Martinez}}, \bibinfo {author} {\bibfnamefont {P.}~\bibnamefont
  {Chan}}, \bibinfo {author} {\bibfnamefont {A.}~\bibnamefont {Rubenok}},
  \bibinfo {author} {\bibfnamefont {C.}~\bibnamefont {John}}, \bibinfo {author}
  {\bibfnamefont {D.}~\bibnamefont {Korchinski}}, \bibinfo {author}
  {\bibfnamefont {C.}~\bibnamefont {Duffin}}, \bibinfo {author} {\bibfnamefont
  {F.}~\bibnamefont {Marsili}}, \bibinfo {author} {\bibfnamefont
  {V.}~\bibnamefont {Verma}}, \bibinfo {author} {\bibfnamefont {M.~D.}\
  \bibnamefont {Shaw}}, \bibinfo {author} {\bibfnamefont {J.~A.}\ \bibnamefont
  {Stern}}, \bibinfo {author} {\bibfnamefont {S.~W.}\ \bibnamefont {Nam}},
  \bibinfo {author} {\bibfnamefont {D.}~\bibnamefont {Oblak}}, \bibinfo
  {author} {\bibfnamefont {Q.}~\bibnamefont {Zhou}}, \bibinfo {author}
  {\bibfnamefont {J.~A.}\ \bibnamefont {Slater}}, \ and\ \bibinfo {author}
  {\bibfnamefont {W.}~\bibnamefont {Tittel}},\ }\href@noop {} {\bibfield
  {journal} {\bibinfo  {journal} {J. Mod. Opt.}\ }\textbf {\bibinfo {volume}
  {62}},\ \bibinfo {pages} {1141} (\bibinfo {year} {2015})}\BibitemShut
  {NoStop}%
\bibitem [{\citenamefont {Takemoto}\ \emph {et~al.}(2015)\citenamefont
  {Takemoto}, \citenamefont {Nambu}, \citenamefont {Miyazawa}, \citenamefont
  {Sakuma}, \citenamefont {Yamamoto}, \citenamefont {Yorozu},\ and\
  \citenamefont {Arakawa}}]{Takemoto15}%
  \BibitemOpen
  \bibfield  {author} {\bibinfo {author} {\bibfnamefont {K.}~\bibnamefont
  {Takemoto}}, \bibinfo {author} {\bibfnamefont {Y.}~\bibnamefont {Nambu}},
  \bibinfo {author} {\bibfnamefont {T.}~\bibnamefont {Miyazawa}}, \bibinfo
  {author} {\bibfnamefont {Y.}~\bibnamefont {Sakuma}}, \bibinfo {author}
  {\bibfnamefont {T.}~\bibnamefont {Yamamoto}}, \bibinfo {author}
  {\bibfnamefont {S.}~\bibnamefont {Yorozu}}, \ and\ \bibinfo {author}
  {\bibfnamefont {Y.}~\bibnamefont {Arakawa}},\ }\href@noop {} {\bibfield
  {journal} {\bibinfo  {journal} {Sci. Rep.}\ }\textbf {\bibinfo {volume}
  {5}},\ \bibinfo {pages} {14383} (\bibinfo {year} {2015})}\BibitemShut
  {NoStop}%
\bibitem [{\citenamefont {Wang}\ \emph {et~al.}(2015)\citenamefont {Wang},
  \citenamefont {Song}, \citenamefont {Yin}, \citenamefont {Wang},
  \citenamefont {Chen}, \citenamefont {Zhang}, \citenamefont {Guo},\ and\
  \citenamefont {Han}}]{Wang15}%
  \BibitemOpen
  \bibfield  {author} {\bibinfo {author} {\bibfnamefont {C.}~\bibnamefont
  {Wang}}, \bibinfo {author} {\bibfnamefont {X.-T.}\ \bibnamefont {Song}},
  \bibinfo {author} {\bibfnamefont {Z.-Q.}\ \bibnamefont {Yin}}, \bibinfo
  {author} {\bibfnamefont {S.}~\bibnamefont {Wang}}, \bibinfo {author}
  {\bibfnamefont {W.}~\bibnamefont {Chen}}, \bibinfo {author} {\bibfnamefont
  {C.-M.}\ \bibnamefont {Zhang}}, \bibinfo {author} {\bibfnamefont {G.-C.}\
  \bibnamefont {Guo}}, \ and\ \bibinfo {author} {\bibfnamefont {Z.-F.}\
  \bibnamefont {Han}},\ }\href@noop {} {\bibfield  {journal} {\bibinfo
  {journal} {Phys. Rev. Lett.}\ }\textbf {\bibinfo {volume} {115}},\ \bibinfo
  {pages} {160502} (\bibinfo {year} {2015})}\BibitemShut {NoStop}%
\bibitem [{\citenamefont {Tang}\ \emph {et~al.}(2016)\citenamefont {Tang},
  \citenamefont {Wei}, \citenamefont {Bedroya}, \citenamefont {Qian},\ and\
  \citenamefont {Lo}}]{Tang16}%
  \BibitemOpen
  \bibfield  {author} {\bibinfo {author} {\bibfnamefont {Z.}~\bibnamefont
  {Tang}}, \bibinfo {author} {\bibfnamefont {K.}~\bibnamefont {Wei}}, \bibinfo
  {author} {\bibfnamefont {O.}~\bibnamefont {Bedroya}}, \bibinfo {author}
  {\bibfnamefont {L.}~\bibnamefont {Qian}}, \ and\ \bibinfo {author}
  {\bibfnamefont {H.-K.}\ \bibnamefont {Lo}},\ }\href@noop {} {\bibfield
  {journal} {\bibinfo  {journal} {Phys. Rev. A}\ }\textbf {\bibinfo {volume}
  {93}},\ \bibinfo {pages} {042308} (\bibinfo {year} {2016})}\BibitemShut
  {NoStop}%
\end{thebibliography}%

\end{document}